\documentclass[aip,jcp,amsmath,amssymb,floatfix,reprint,citeautoscript,longbibliography,noeprint]{revtex4-2}

\usepackage[utf8]{inputenc}
\usepackage[T1]{fontenc}
\usepackage{lmodern}
\usepackage[version=4]{mhchem}
\usepackage{graphicx}
\usepackage{wrapfig}\usepackage[colorlinks,allcolors=black,citecolor=blue,urlcolor=blue]{hyperref}
\usepackage[mathlines,displaymath]{lineno}
\usepackage{placeins}
\usepackage{booktabs}
\usepackage{tikz}
\usetikzlibrary{decorations.pathreplacing,calligraphy}   
\usetikzlibrary{shapes.misc, positioning}
\usepackage{tabularx}
\usepackage{xcolor}

\DeclareUnicodeCharacter{2212}{-}
\graphicspath{{figures/}}

\begin{document}
\def\mytitle{%
Clay Edges Are Dynamic Proton-conducting Networks Modulated by Structure and pH
}
\title{\mytitle}
\author{Yixuan Feng}%
\affiliation{%
State Key Laboratory of Hydro-science and Engineering, Department of Hydraulic Engineering, Tsinghua University, Beijing 100084, China
}
\author{Xavier R. Advincula}%
\affiliation{%
Yusuf Hamied Department of Chemistry, University of Cambridge, Lensfield Road, Cambridge, CB2 1EW, UK
}
\affiliation{%
Cavendish Laboratory, Department of Physics, University of Cambridge, Cambridge, CB3 0HE, UK
}
\affiliation{%
Lennard-Jones Centre, University of Cambridge, Trinity Ln, Cambridge, CB2 1TN, UK
}
\author{Hongwei Fang}%
\email{fanghw@tsinghua.edu.cn}
\affiliation{%
State Key Laboratory of Hydro-science and Engineering, Department of Hydraulic Engineering, Tsinghua University, Beijing 100084, China
}

\author{Christoph Schran}%
\email{cs2121@cam.ac.uk}
\affiliation{%
Cavendish Laboratory, Department of Physics, University of Cambridge, Cambridge, CB3 0HE, UK
}
\affiliation{%
Lennard-Jones Centre, University of Cambridge, Trinity Ln, Cambridge, CB2 1TN, UK
}

\date{\today}
\begin{abstract}
Montmorillonite, a ubiquitous clay mineral, plays a vital role in geochemical and environmental processes due to its chemically complex edge surfaces.
However, the molecular-scale acid–base reactivity of these interfaces remains poorly understood due to the limitations of both experimental resolution and conventional simulations.
Here, we employ machine learning potentials with first-principles accuracy to perform nanosecond-scale molecular dynamics simulations of montmorillonite nanoparticles across a range of pH.
Our results reveal clear amphoteric behavior: edge sites undergo protonation in acidic environments and deprotonation in basic conditions.
Even at neutral pH, spontaneous and directional proton transfer events are common, proceeding via both direct and solvent-mediated pathways.
These findings demonstrate that montmorillonite edges are not static arrays of hydroxyl groups but dynamic, proton-conducting networks whose reactivity is modulated by local structure and solution conditions.
This work offers a molecular-level framework for understanding proton transport and buffering in clay–water systems, with broad implications for catalysis, ion exchange, and environmental remediation.
\end{abstract}

{\maketitle}
\renewcommand{\tocname}{\vspace*{-2em}}



\section{Introduction}
Clay minerals are ubiquitous in terrestrial and marine environments and play an essential role in a wide range of geochemical and environmental processes, including nutrient cycling, metal mobilization, ice nucleation, and sediment transformation\cite{doi:10.1073/pnas.1800141115,
doi:10.1038/s41586-019-1280-6,
doi:10.1038/s41467-022-31004-0,
doi:10.1021/acs.jpclett.6b01013,
doi:10.1038/s43017-022-00301-z}.
In technological applications, clay-based materials are widely employed in pollution control, water purification, and radioactive waste containment owing to their exceptional cation exchange capacity and adsorption performance\cite{doi:10.1346/CCMN.2013.0610601,doi:10.1016/j.cej.2016.09.029,doi:10.1016/B978-0-08-100027-4.00002-4}.
To fully elucidate the role of clay minerals in these processes and to optimize their functional applications, it is crucial to develop a detailed understanding of their adsorption capabilities\cite{doi:10.1021/acs.jpcc.1c05995}, dissolution and growth kinetics\cite{doi:10.2138/am-2001-0404}, and aggregation behavior\cite{doi:10.1073/pnas.1621186114}.
Underlying all of these phenomena is the surface chemistry of clay minerals, particularly their acid–base properties and surface proton dynamics.

Among these minerals, montmorillonite---a swelling smectite with a layered aluminosilicate structure---has emerged as one of the most extensively studied clays due to its high specific surface area, thermal and mechanical stability, and intrinsic structural charge heterogeneity\cite{doi:10.1038/s41467-020-17801-5,doi:10.2138/am-2022-8834}.
These properties render montmorillonite a highly effective cation sorbent with broad applications in heavy metal removal and contaminant immobilization.
However, despite decades of research, a fundamental molecular-level understanding of montmorillonite surface reactivity remains elusive.
The basal surfaces, dominated by siloxane sheets, are chemically inert and largely pH-independent\cite{doi:10.1073/pnas.96.7.3358}.
In contrast, edge surfaces expose amphoteric hydroxyl groups that, although less abundant, are central to proton exchange, metal complexation, and mineral dissolution, particularly under variable pH conditions\cite{doi:10.1016/j.clay.2004.01.001,doi:10.1016/j.clay.2006.05.009,doi:10.1016/j.chemosphere.2004.09.023}.
The structural heterogeneity and chemical complexity of these edge sites, particularly in environmentally relevant aqueous systems, continue to present major challenges for direct characterization, thus hindering mechanistic understanding.

Extensive experimental efforts have been made to probe the acid–base reactivity of montmorillonite surfaces, particularly through acid–base titration\cite{doi:10.1016/j.jcis.2005.03.060,doi:10.1016/S0169-7722(97)00008-9,doi:10.1016/j.jcis.2006.04.081}.
However, despite careful sample pretreatment and modeling efforts, it remains difficult to independently determine the density and intrinsic acidity constants of individual surface functional groups.
Instead, surface protonation is often inferred from macroscopic titration curves using surface complexation models (SCMs), which rely on simplified assumptions and cannot unambiguously resolve the contributions of specific edge sites\cite{doi:10.1016/j.jcis.2007.03.062,doi:10.2475/05.2013.01,doi:10.1021/acs.est.6b04677}.
The structural and chemical heterogeneity of montmorillonite edge surfaces, encompassing the diversity of Si- and Al-associated hydroxyl groups as well as the effects of isomorphic substitutions, further complicates experimental characterization and molecular modelling efforts\cite{doi:10.1016/S1572-4352(05)01003-2}.

To overcome the limitations of experimental approaches in resolving surface protonation at the atomic level, \textit{ab initio} molecular dynamics (AIMD) simulations have been widely employed to investigate the acid--base behavior of clay mineral edges\cite{doi:10.1016/B978-0-08-102432-4.00003-2}.
Multiple studies have calculated the intrinsic acidity constants (\textit{p}K\textsubscript{a}) of individual hydroxyl groups on montmorillonite, kaolinite, and pyrophyllite edge surfaces, demonstrating how site-specific chemistry is modulated by structural factors such as coordination state and isomorphic substitution\cite{doi:10.1016/j.gca.2013.04.008, doi:10.1016/j.gca.2014.05.044, doi:10.1016/j.gca.2012.07.010, doi:10.1016/j.gca.2015.07.015}.
These simulations consistently confirm the amphoteric nature of edge sites, and suggest that substitutions in both tetrahedral and octahedral sheets can significantly alter local acid strength, often elevating the \textit{p}K\textsubscript{a} and stabilizing the protonated state.
Beyond \textit{p}K\textsubscript{a} estimation, AIMD studies have also explored spontaneous proton transfer (PT) dynamics and hydration structures at fully solvated clay edges\cite{doi:10.1016/j.gca.2015.07.013, doi:10.1016/j.gca.2006.11.026}.
These works reveal that surface reactivity strongly depends on crystallographic orientation and interfacial hydrogen-bond networks.
For example, the (010) edge of pyrophyllite has been shown to support spontaneous proton exchange between adjacent hydroxyl groups via water-mediated pathways, while the (110) surface remains largely inert under comparable conditions.
Interlayer and micropore water can further stabilize charged species, emphasizing the need to consider the full solvation environment. 

Despite these advances, current simulations remain limited by spatial and temporal scales.
Most AIMD studies probe only isolated surface reactions over picosecond timescales and predominantly under neutral pH conditions, offering limited access to rare events, correlated proton dynamics, or evolving charge states across extended clay edges.
Direct molecular-level observations of acid–base reactivity in fully hydrated clay nanoparticles, which are the relevant reactive units in natural and engineered settings, are essentially absent, leaving key questions unresolved.
In particular, the interplay between amphoteric surface chemistry and proton transfer has yet to be established.
Addressing this gap requires modeling strategies that retain quantum-level accuracy for reactive edge groups while enabling nanosecond-scale sampling of realistic, solvated mineral surfaces across a range of pH.

Recent developments in machine learning potentials (MLPs) have provided a powerful and efficient framework for simulating chemically complex clay–water interface systems.
By learning the potential energy surface from first-principles data, MLPs dramatically reduce computational cost while preserving \textit{ab initio} accuracy, thereby enabling nanosecond-scale simulations of large, solvated mineral surfaces~\cite{doi:10.1073/pnas.2110077118}.
This approach has already proven effective in modeling structural and mechanical properties of phyllosilicates such as kaolinite~\cite{doi:10.1016/j.clay.2022.106596, doi:10.1063/5.0152361} and pyrophyllite~\cite{doi:10.1021/acs.jpca.5c00406}, yielding results consistent with both density functional theory (DFT) and experimental data.
Beyond static structure, MLPs have been used to investigate dynamic interfacial phenomena such as water layering, hydrogen-bond dynamics, and ion exchange, achieving temporal and spatial resolution beyond the reach of conventional AIMD simulations~\cite{doi:10.1021/acs.jpcc.4c03288, doi:10.1063/5.0217720}.
%
Furthermore, MLPs retain a fully reactive description at near quantum-level accuracy.
Although not yet applied to proton transport in clays, recent studies have successfully captured acid–base reactions and proton transport at graphitic and oxide surfaces, as well as within confined aqueous environments~\cite{doi:10.1038/s41557-024-01593-y,doi:10.1021/acs.jpclett.6b01448, doi:10.1021/acs.jpcc.4c05857, doi:10.1021/acsnano.5c02053}, offering detailed insights into bond-breaking, charge delocalization, and solvation effects.
These advances show that MLPs are uniquely suited to bridge the current knowledge gap in clay edge chemistry by capturing both the structural heterogeneity and the dynamic proton exchange processes at the mineral–water interface over extended timescales.
In this context, MLP-based simulations offer a promising path forward to resolve the coupled mechanisms of surface acid–base reactivity and PT in realistic clay nanoparticle systems.

In this study, we leverage the accuracy and efficiency of MLPs to investigate the surface chemistry of montmorillonite in aqueous environments.
%
Our simulations reveal the amphoteric nature of edge surfaces, capturing the protonation and deprotonation behavior of structurally diverse reactive sites.
By resolving the free energy landscapes of representative PT events, we quantify the thermodynamics of various exchange pathways and gain mechanistic insight into site-specific reactivity.
Furthermore, we examine how isomorphic substitution modulates local acid–base behavior by elevating the free energy barriers for deprotonation.
Ultimately, our findings elucidate fundamental mechanisms of acid–base reactivity at clay–water interfaces, offering transferable insights into mineral–solution interactions in geochemistry, environmental remediation, and heterogeneous catalysis.
\label{sec:intro}
\begin{figure*}
    \centering{}
    \includegraphics[width=\textwidth]{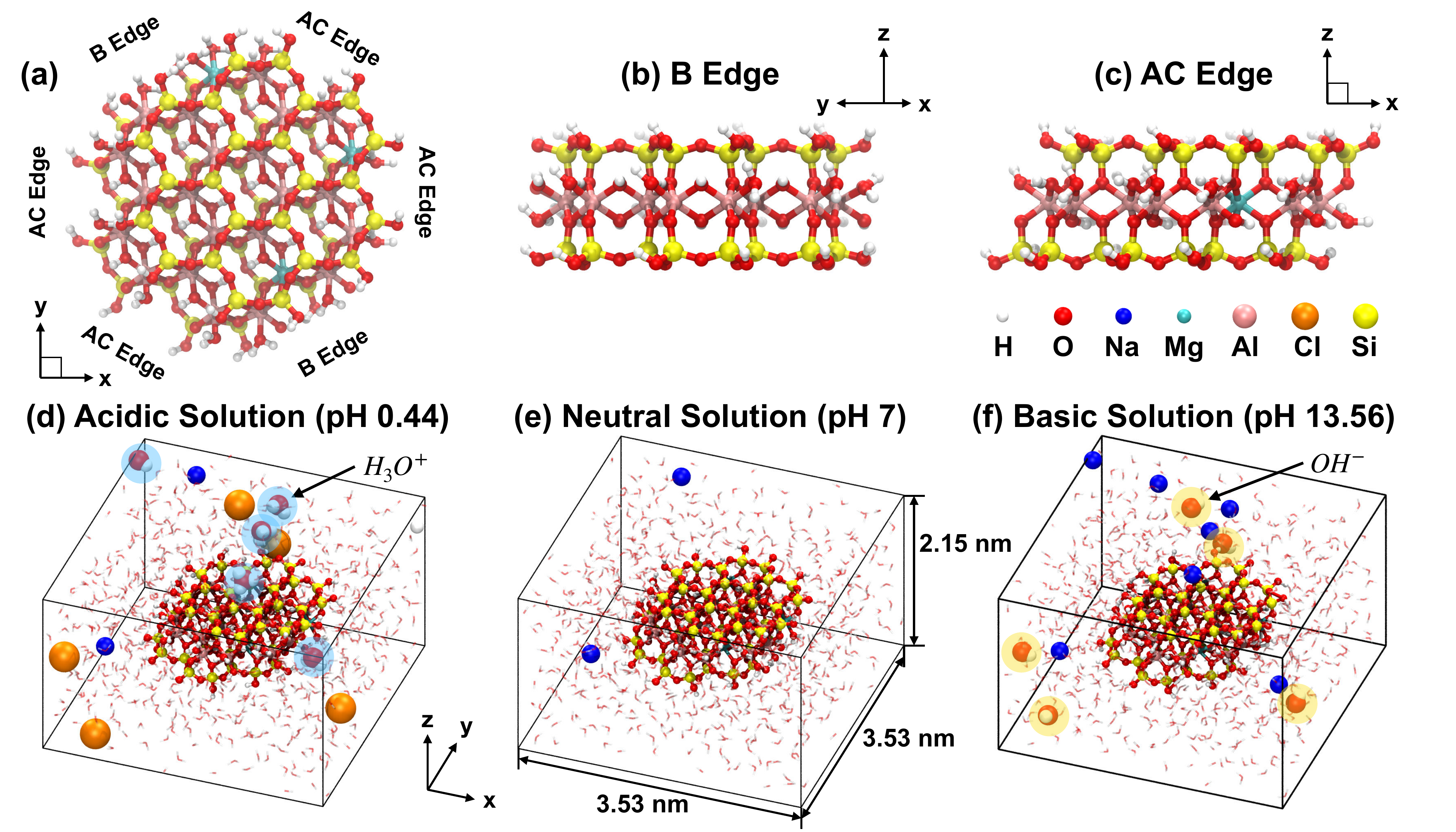}
    \caption{
        \textbf{Snapshots of the simulation system comprising a montmorillonite nanoparticle surrounded by aqueous solution.}
	(a) Top view of the clay nanoparticle.
	(b) Side view perpendicular to the B edge.
	(c) Side view perpendicular to the AC edge.
        (d–f) Representative snapshots of the nanoparticle in aqueous solution under different pH conditions: (d) acidic (pH~0.44), (e) neutral (pH~7), and (f) basic (pH~13.56).
        Coordinate axes are provided to indicate the viewing directions.
    }
    \label{fig:figure1}
\end{figure*}

\section{Results and Discussions}

%
Montmorillonite is a 2:1 dioctahedral phyllosilicate clay mineral characterized by isomorphic substitution of Al by Mg in its octahedral sheet, described by a general formula of M$^+_y$[Si$_8$][Al$_{4-y}$Mg$_y$]O$_{20}$(OH)$_4$.
Given the dominant role of edge sites in governing acid–base reactivity of clays, as established in prior experimental and theoretical studies, our simulations focused on capturing the structure and chemical dynamics of montmorillonite edge surfaces in aqueous environments.
Informed by atomic force microscopy (AFM) measurements and molecular dynamics simulations, the (110) and (010) edge surfaces---commonly referred to as the AC and B edges~\cite{doi:10.1346/CCMN.1988.0360207} ---are identified as the most frequently exposed terminations, accounting for approximately 60\% and 20\% of the total edge area, respectively~\cite{doi:10.1016/j.clay.2020.105442,doi:10.1021/jp053874m}.
To reflect this distribution, we constructed hexagonal-shaped montmorillonite nanoparticles comprising four AC and two B edges (Fig.~\ref{fig:figure1}a).
While ideal hexagonal symmetry is rarely observed in natural specimens, this construction enables simultaneous exploration of diverse edge types and their potential cooperative interactions within a single simulation setup.
The atomic structures of the B and AC edge terminations are shown in Figures~\ref{fig:figure1}b and \ref{fig:figure1}c.
All broken bonds at the edges were chemically terminated with hydroxyl or hydrogen groups to maintain local charge neutrality.
To investigate the role of compositional disorder, three Mg-for-Al isomorphic substitutions were randomly introduced into each nanoparticle, and three distinct substitution patterns were considered (Mont.1, Mont.2, and Mont.3; see Supplementary Section~S1).
To examine the pH-dependent reactivity of edge sites, each nanoparticle was simulated in acidic (pH~0.44), neutral (pH~7), and basic (pH~13.56) aqueous environments (Figures~\ref{fig:figure1}d–f), including respective counter ions to retain charge neutrality.
To enable a detailed exploration of chemical reactivity in these systems, a MLP was developed using the MACE framework~\cite{doi:10.48550/arXiv.2206.07697} and trained on revPBE-D3 reference data (see Methods).
The training dataset encompassed bulk aqueous solutions and clay–water interface structures under a range of pH conditions, allowing the model to accurately reproduce energies and forces across diverse chemical environments.
The resulting potential reliably captures interfacial energetics, clay lattice properties, and the structure of aqueous phases (see Supplementary Section~S1-S2).
Nanosecond-scale molecular dynamics simulations using this MLP enabled us to examine the acid–base reactivity of montmorillonite edge surfaces in various aqueous environments, capturing their dynamic protonation states and associated PT behavior under chemically realistic conditions.

\subsection{Amphoteric behavior of edge surfaces}
\begin{figure*}
    \centering{}
    \includegraphics[width=\textwidth]{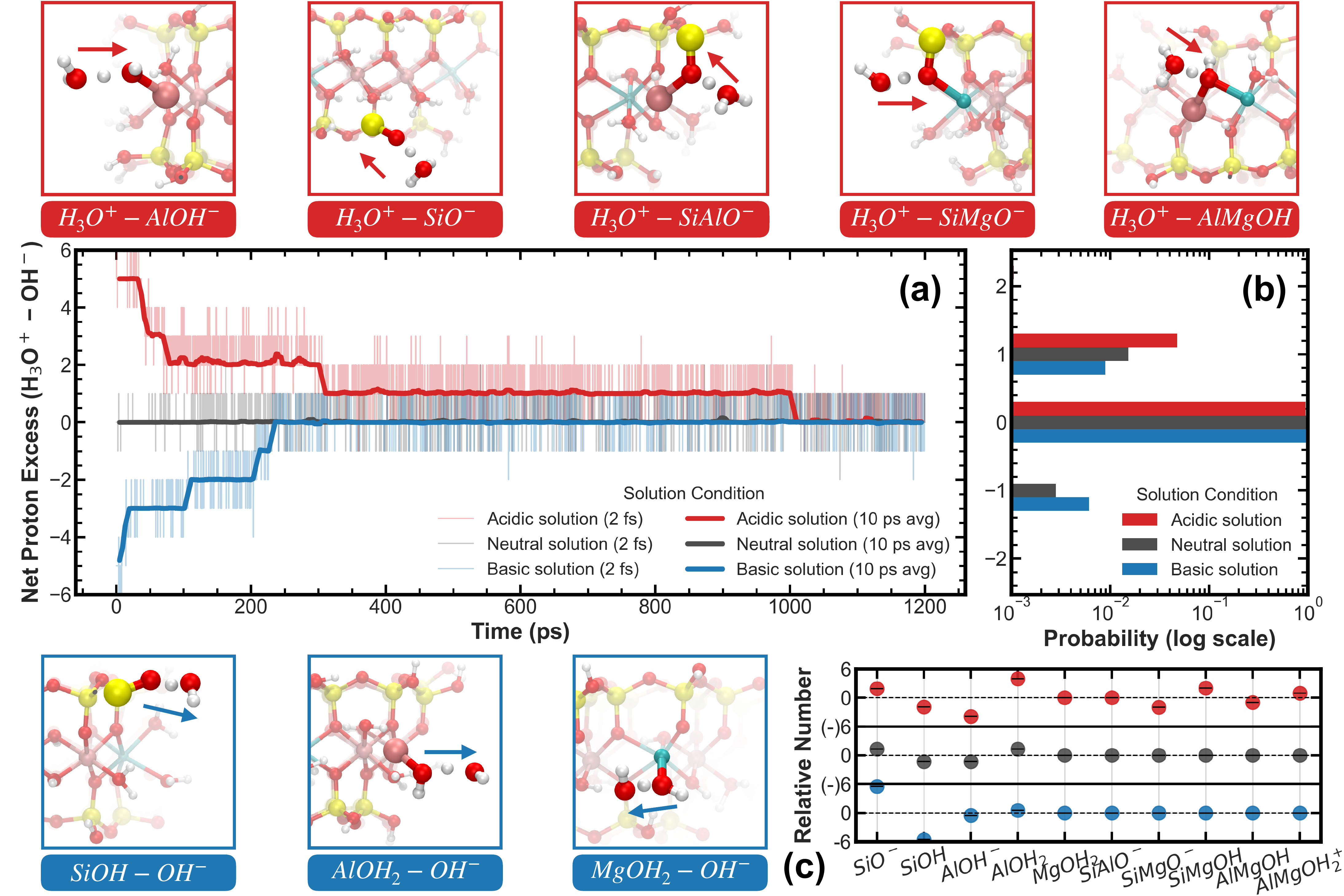}
    \caption{
        \textbf{Protonation and deprotonation behavior of the montmorillonite nanoparticle in different aqueous environments.}
        (a) Time evolution of the net proton excess in the aqueous phase (defined as the number of hydronium ions minus hydroxide ions) over 1200 ps.
        Transparent curves represent data sampled every 2 fs; solid lines denote a 10 ps average.
        (b) Probability distribution of the net proton excess in the aqueous phase during the final 200 ps.
        (c) Relative abundance of various surface functional groups during the final 200 ps, compared to their initial populations.
        In all panels, colors denote the solution pH: red for acidic, black for neutral, and blue for basic conditions.
        Top panel: Surface groups that gain protons from hydronium ions under acidic conditions.
        Bottom panel: Surface groups that donate protons to hydroxide ions under basic conditions.
        Arrows indicate the direction of proton transfer, with representative surface reactions labeled below each group pair.
    }
    \label{fig:figure2}
\end{figure*}

To explore the acid--base reactivity of montmorillonite edge surfaces, we first examined the dynamic protonation behavior of nanoparticles under acidic, neutral, and basic conditions.
Fig.~\ref{fig:figure2} presents representative results for one of the studied montmorillonite structures; analogous behavior was observed for the other two nanoparticle configurations (see Supplementary Section~S3).
In Fig.~\ref{fig:figure2}a, the solution-phase net proton excess, defined as the number of hydronium ions minus hydroxide ions, starts at about +5 under acidic conditions and about −5 under basic conditions.
As the simulation proceeds, the mean absolute value in both cases gradually decreases and approaches zero, indicating progressive proton uptake by the nanoparticle in acidic environments and proton release in basic environments.
This opposite pH-dependent trend demonstrates the amphoteric character of the nanoparticle.
This is consistent with previous \textit{ab initio} simulations, which inferred amphoteric character from a limited number of proton transfer events at specific edge sites under neutral conditions~\cite{doi:10.1016/j.gca.2015.07.013}. 
In contrast, our simulations evaluate protonation behavior across a range of pH environments, allowing a more comprehensive and chemically direct assessment of the ampotheric nature of montmorrilonite based on the net acid--base response of the particle.
This macroscopic amphoteric trend is also consistent with earlier titration and coagulation experiments on montmorillonite suspensions, which suggested pH-dependent charge development at edge sites~\cite{doi:10.1016/j.clay.2004.01.001}.
While this qualitative behavior is robust across structural variations, the kinetics of protonation and deprotonation differ significantly: the average time required to lose five protons in basic solution was approximately $179 \pm 46$~ps, whereas acquiring five protons in acidic solution occurred no earlier tha
634~ps and was not completed within 1.2~ns in some cases (see Supplementary Section~S3).

Trajectory analysis revealed distinct sets of reactive surface groups under different conditions.
In acidic environments, protonation occurred primarily at negatively charged or undercoordinated sites, including \ce{-AlOH^-}, \ce{-SiO^-}, \ce{-SiAlO^-}, \ce{-SiMgO^-}, and \ce{-AlMgOH}.
In basic environments, deprotonation involved groups such as \ce{-SiOH}, \ce{-AlOH2}, and \ce{-MgOH2} (Fig.~\ref{fig:figure2}, top and bottom panels).
These differences reflect not only the intrinsic acid--base properties of the functional groups but also their abundance and spatial distribution. Sites reactive toward hydroxide ions are more prevalent and broadly distributed across both AC and B edges.
In contrast, groups readily protonated by hydronium ions are more localized, primarily occurring along B edges and AC edges that contain isomorphic substitutions.
Among these, the \ce{-AlMgOH} group is of particular interest.
Although rarely considered in previous structural models, this site at the intersection of two AC edges frequently undergoes protonation in our simulations.
However, its reactivity appears contingent upon the presence of Mg substitution at one of the adjacent Al sites; systems lacking this substitution show no clear protonation of the corresponding \ce{-AlAlOH} group.
These findings highlight the critical role of isomorphic substitution in modulating local acid--base behavior, later analysed in more detail.
The proton affinities of other active sites, including \ce{-AlOH^-}, \ce{-SiAlO^-}, and \ce{-SiMgO^-}, are well supported by prior computational studies~\cite{doi:10.1016/j.gca.2011.12.009, doi:10.1021/jp053874m}.
While these studies predicted high proton affinity based on free energy calculations or Fukui function analysis, our simulations directly capture the protonation of these sites by hydronium ions under acidic conditions, providing dynamic, atomistic confirmation of their Brønsted basicity.

Fig.~\ref{fig:figure2}c quantifies the relative population changes of key surface groups in the final 200~ps.
In acidic conditions, protonated species such as \ce{-AlOH_2}, \ce{-SiMgOH}, and \ce{-AlMgOH_2^+} became more abundant, forming not only via protonation by hydronium ions but also through PT from \ce{-SiOH}.
Conversely, in basic solutions, \ce{-SiOH} groups transferred protons both to hydroxide ions in solution and to nearby \ce{-AlOH^-} groups.
These observations highlight that proton exchange occurs not only between surface sites and the bulk solution but also among surface groups themselves. 
In neutral solutions, the decrease in \ce{-SiOH} and the increase in \ce{-AlOH_2} suggest that spontaneous PT occurs between these two surface groups.
While these events are consistent with previous simulation studies that reported proton transfer between \ce{-SiOH} and \ce{-AlOH^-} groups at clay edges~\cite{doi:10.1016/j.gca.2015.07.013, doi:10.1016/j.gca.2006.11.026}, those works captured only isolated, short-lived PT events over limited simulation timescales. 
For instance, Churakov~\cite{doi:10.1016/j.gca.2006.11.026} observed a reversible PT between these sites within several picoseconds, while Greathouse et al.~\cite{doi:10.1016/j.gca.2015.07.013} reported unidirectional transfer without reversal under specific hydration environments. 
In contrast, our long-timescale simulations reveal a statistically favored net deprotonation of \ce{-SiOH} and protonation of \ce{-AlOH^-} groups under neutral conditions, suggesting that this directional PT process is not merely transient, but reflects a thermodynamically preferred configuration.

This spontaneous PT is accompanied by persistent fluctuations in the net proton excess of the aqueous phase, indicating that water molecules, in the form of hydrated hydronium ions and hydroxide ions, play a role in these PT events.
Although the probability of these events occurring is generally low,
they are indicative of the dynamic nature of proton exchange on the surface (Fig.~\ref{fig:figure2}b).
These findings highlight the importance of understanding the site-specific mechanisms underlying PT, which are crucial for interpreting surface reactivity and charge dynamics.

\subsection{Rich proton transfer dynamics under neutral conditions}
\begin{figure*}
    \centering{}
    \includegraphics[width=\textwidth]{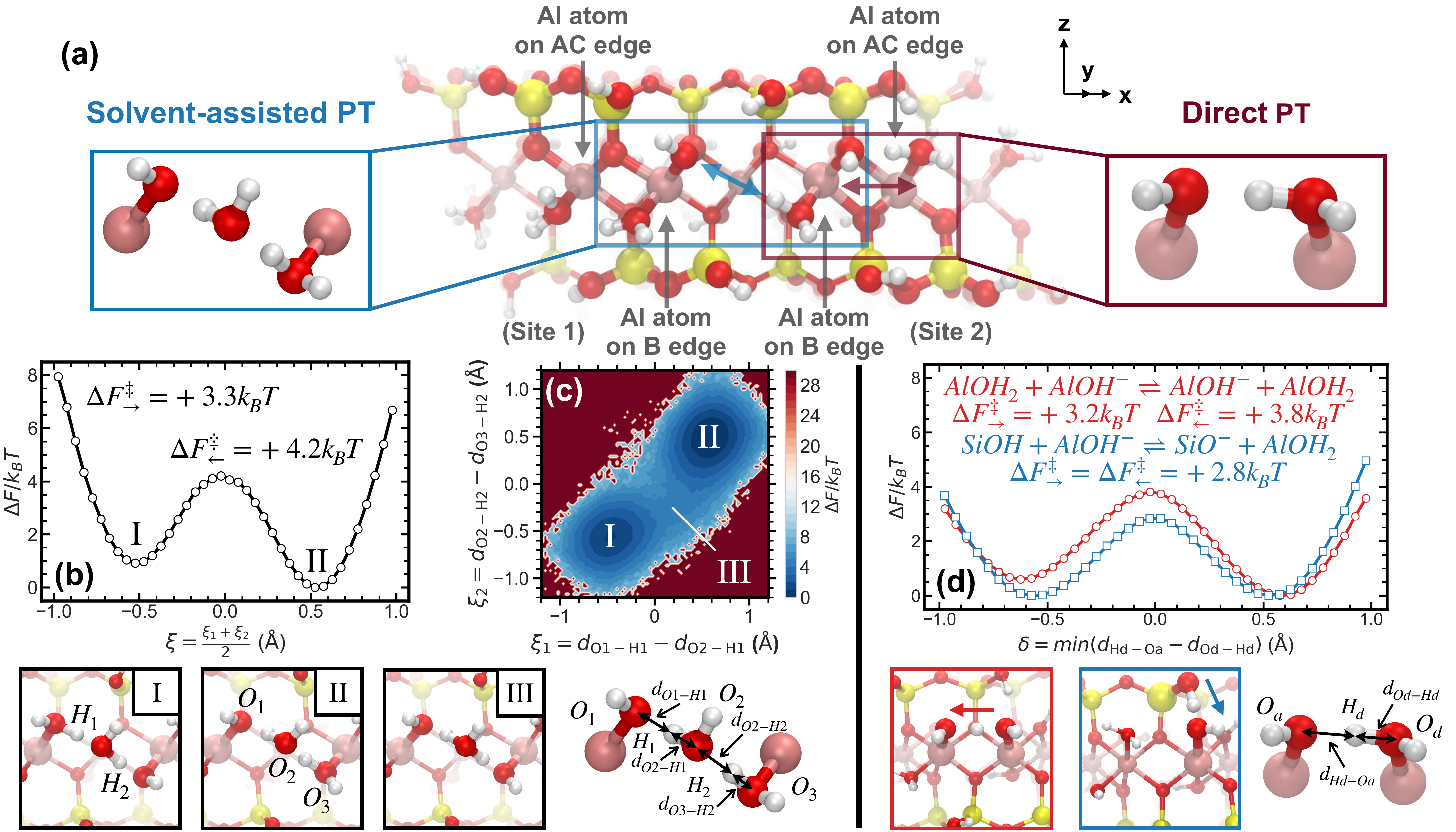}
    \caption{
        \textbf{Proton transfer mechanisms occurring spontaneously on the montmorillonite nanoparticle  edge surface under neutral aqueous conditions.} %
	(a) Frontal view of the B edge, showing two distinct surface environments relevant to proton transfer. Site~1 consists of an \(\mathrm{-AlOH^-}\) group adjacent to a \(\mathrm{-SiAlO^-}\) group and primarily undergoes solvent-assisted proton transfer. Site~2 features an \(\mathrm{-AlOH^-}\) group adjacent to an \(\mathrm{-AlOH_2}\) group and predominantly supports direct proton transfer. Arrows indicate the dominant PT directions at each site. %
    \textbf{Left panel (site 1, solvent-assisted PT):} %
    (b) Projected free energy profile as a function of the averaged coordinate \(\xi = (\xi_1 + \xi_2)/2\). %
    (c) Free energy surface for a solvent-assisted proton transfer process between \(\mathrm{-AlOH^-}\) and \(\mathrm{-AlOH_2}\) groups as a function of two collective variables, \(\xi_1 = d_{\mathrm{O1}-\mathrm{H1}} - d_{\mathrm{O2}-\mathrm{H1}}\) and \(\xi_2 = d_{\mathrm{O2}-\mathrm{H2}} - d_{\mathrm{O3}-\mathrm{H2}}\), in units of \AA. %
    Representative configurations (labeled \textit{I}, \textit{II} and \textit{III}) corresponding to key points on the free energy surface are shown below panels (b) and (c). %
    \textbf{Right panel (site 2, chain-like direct PT):} %
    (d) Free energy profiles for two chain-like proton transfer reactions. %
    The red curve with circles corresponds to the reaction \(\mathrm{AlOH_2 + AlOH^- \rightleftharpoons AlOH^- + AlOH_2}\), while the blue curve with squares corresponds to \(\mathrm{SiOH + AlOH^- \rightleftharpoons SiO^- + AlOH_2}\). %
    Representative structures are shown adjacent to the curves, with arrows indicating the direction of proton transfer.
    }
    \label{fig:figure3}
\end{figure*}

To further elucidate the spontaneous PT observed in neutral solutions, we examined the structural origins and pathways underlying this behavior across all three nanoparticle models.
Trajectory analysis revealed that the dynamic reactivity of \ce{-AlOH^-} groups located on the B edge plays a central role.
These sites consistently exhibit a tendency to acquire protons from neighboring functional groups, a behavior previously reported on similar aluminosilicate edges~\cite{doi:10.1016/j.gca.2006.11.026}.
In our hexagonal nanoparticle model, each B edge contains two aluminum atoms, forming two \ce{-Al(OH)(OH_2)} terminal motifs at junctions with two symmetrically opposite AC edges.
Due to the mirror symmetry of these adjacent AC edges---one corresponding to the (110) surface and the other to the ($\bar{1}$10) surface---the two resulting \ce{-AlOH^-} sites exhibit distinct local environments and are denoted as site~1 and site~2 (Fig.~\ref{fig:figure3}a): site~1 lies adjacent to a \ce{-SiAlO^-} group, while site~2 neighbors a \ce{-AlOH_2} group.

These local differences give rise to distinct PT mechanisms.
At site~1, the spatial separation between the \ce{-AlOH^-} and nearby proton donors hinders direct transfer.
Instead, protonation proceeds through a solvent-assisted mechanism mediated by a bridging water molecule that links the \ce{-AlOH_2} site on the same B edge.
This solvent-assisted PT pathway was observed consistently across multiple surface environments, with hundreds of events recorded over our 1.2~ns simulations, indicating its dynamic relevance and justifying further free energy analysis.
To quantify the free energy barrier associated with this solvent-assisted PT, we first computed a one-dimensional proton transfer free energy landscape (PTFEL) along the averaged reaction coordinate $\xi = (\xi_1 + \xi_2)/2$, where each $\xi_i$ reflects the position of a transferring proton relative to its donor and acceptor oxygen atoms (see Methods).
The resulting profile shows that the free energy barrier for protonation of \ce{-AlOH^-} ($\Delta F^{\ddagger}_{\leftarrow} = +4.2\,k_BT$) is slightly higher than that for deprotonation ($\Delta F^{\ddagger}_{\rightarrow} = +3.3\,k_BT$), indicating a mild energetic bias toward the deprotonated state (Fig.~\ref{fig:figure3}b).
To further resolve the transition pathway, we constructed a two-dimensional PTFEL using the full set of collective variables $(\xi_1, \xi_2)$ (Fig.~\ref{fig:figure3}c), highlighting three key configurations.
States~I and~II correspond to the final and initial protonation states, while state~III represents a hydronium-like intermediate along the dominant low-energy pathway.
Although not labeled in the figure, hydroxide-like configurations with less favourable free energy are also accessible and are expected to contribute to the net proton fluctuations observed in Fig.~\ref{fig:figure2}.
Overall, these features suggest that PT proceeds through multiple reversible fluctuations rather than a single well-defined hopping event.

At site~2, the neighboring \ce{-AlOH_2} group enables a direct transfer pathway.
Here, protonation of the B edge \ce{-AlOH^-} is facilitated by deprotonation of the AC edge \ce{-AlOH_2}, which in turn can receive a proton from a nearby \ce{-SiOH}, forming a chain-like sequence of transfers.
The corresponding PTFEL was constructed using a one-dimensional reaction coordinate $\delta$~\cite{doi:10.1038/nature00797,doi:10.1021/acs.jpclett.7b00358}, with sign assigned according to the direction of the chemical transformation (Fig.~\ref{fig:figure3}d).
The barrier for protonation of the B edge site ($\Delta F^{\ddagger}_{\rightarrow} = +3.2\,k_BT$) is lower than that of the reverse process ($\Delta F^{\ddagger}_{\leftarrow} = +3.8\,k_BT$), indicating a moderate thermodynamic preference for the protonated state.
Following this step, the AC edge \ce{-AlOH_2} group, after donating a proton, can regain their proton from a neighboring \ce{-SiOH} group.
The corresponding PT between \ce{-AlOH^-} and \ce{-SiOH} exhibits nearly symmetric barriers of approximately $2.8\,k_BT$, suggesting a dynamic equilibrium between these two surface groups.
Both the direct PT between \ce{-AlOH^-} and \ce{-AlOH_2} and the subsequent transfer involving \ce{-SiOH} occurred frequently, with hundreds of events recorded over the 1.2~ns simulation window.
The latter pathway was observed approximately two to three times more often, reinforcing its role as the primary channel for dynamic proton exchange at the interface.
This balance facilitates frequent bidirectional PTs among the B edge and AC edge groups, leading to transient surface charge fluctuations.

Together, these two pathways explain the distinct features observed in neutral conditions: the solvent-assisted transfer at site~1 accounts for the persistent net proton excess fluctuations (Fig.~\ref{fig:figure2}b), while the direct proton hopping at site~2 explains the changes in surface group populations (Fig.~\ref{fig:figure2}c).
These mechanisms were consistently observed across all three nanoparticle models, highlighting their general significance.
However, they are not the only PT routes accessible to B edge \ce{-AlOH^-} groups.
Additional transfer events involving neighboring \ce{-SiOH} groups were also detected, proceeding via either direct or solvent-assisted mechanisms (see Supplementary Fig.~S9).
These alternative pathways, while mechanistically plausible, occurred much less frequently in our simulations, with only a handful to a dozen events observed over the 1.2~ns trajectories, suggesting they play a minor role in the overall interfacial proton dynamics.

\subsection{Changes in pH facilitate more diverse proton transfer}
\begin{figure}
    \centering{}
    \includegraphics[width=0.48\textwidth]{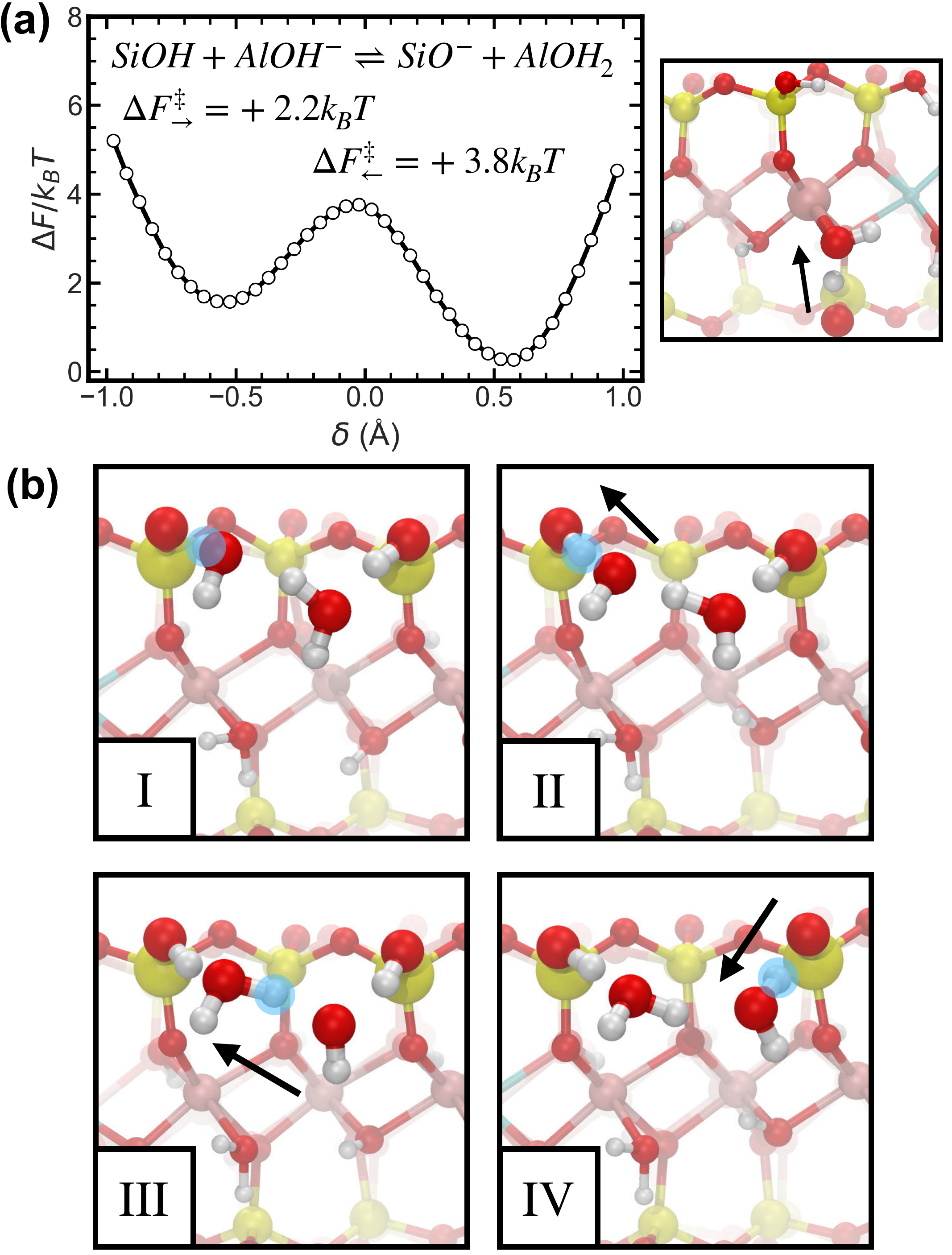}
    \caption{
        \textbf{Proton transfer processes on the AC edge of the montmorillonite nanoparticle in basic solution.}
        (a) Free energy profile for a direct proton transfer reaction on the AC edge, involving surface groups not adjacent to B edge junctions: \(\mathrm{SiOH + AlOH^- \rightleftharpoons SiO^- + AlOH_2}\).
        (b) Representative snapshots illustrating water-mediated multi-step proton transfer between \(\mathrm{-SiOH}\) and \(\mathrm{-SiO^-}\) groups. Four sequential structures (I–IV) are shown, with arrows indicating the direction of proton migration through bridging water molecules.
    	\label{fig:figure4}
    }
\end{figure}


While the preceding sections focused on neutral conditions, PT behaviors in acidic and basic solutions also exhibit distinct mechanistic features.
Across all systems, we observed a pronounced dependence of PT frequency on solution pH. 
In acidic conditions, the number of PT events was significantly reduced compared to neutral solution, often by more than half and in some cases by an order of magnitude.
This suppression arises because early protonation of key surface sites stabilizes them in their conjugate acid forms, thereby preventing further exchange.
In contrast, basic conditions yield a more heterogeneous response: for some sites, PT frequency increased due to enhanced deprotonation and interaction with hydroxide ions, while for others it declined, depending sensitively on the local structure and accessibility of reactive partners.

These general trends manifest in system-specific ways, as illustrated by the following representative observations under acidic and basic conditions. 
In acidic solutions, early protonation of surface groups limits subsequent PT activity. 
For instance, in the Mont.1 and Mont.3 systems, \ce{-SiOH} groups remained fully protonated during the final 200~ps (see Supplementary Fig.~S6).
In contrast, basic environments promote deprotonation events that trigger more diverse PT processes.
Notably, in addition to the B-edge mechanisms discussed above, B-edge \ce{-AlOH^-} groups in basic solution were frequently observed to be involved in PT with nearby \ce{-SiOH} groups, proceeding through both direct and solvent-assisted pathways.
Although these interactions consistently involve the same chemical species, i.e., \ce{-AlOH^-} and \ce{-SiOH}, structural analysis shows that their local environments---determined by edge geometry and relative positioning---give rise to three distinct classes of PT pathways.
In some cases, the directionality and energy barrier of proton exchange depend critically on whether the transfer proceeds directly or via a bridging water molecule (see Supplementary Fig.~S9).

AC edge sites also exhibit distinct reactivity patterns.
Although relatively stable under acidic and neutral conditions, surface groups such as \ce{-SiOH} or \ce{-AlOH_2} can undergo deprotonation through reactions with hydroxide ions in basic solution, which in turn may initiate subsequent proton transfer cascades.
Two representative mechanisms are shown in Fig.~\ref{fig:figure4} and Fig.~S10.
The first is a direct PT reaction involving \ce{-SiOH} and \ce{-AlOH^-} groups, leading to the formation of \ce{-SiO^-} and \ce{-AlOH_2}.
While either configuration may serve as the initial state in simulations, the associated one-dimensional PTFEL was constructed along the forward reaction pathway, \ce{-SiOH + AlOH^- -> SiO^- + AlOH_2}, using a signed reaction coordinate $\delta$.
The resulting free energy profile shows a lower forward barrier ($\Delta F^{\ddagger}_{\rightarrow} = +2.2\,k_BT$) compared to the reverse ($\Delta F^{\ddagger}_{\leftarrow} = +3.8\,k_BT$), indicating a thermodynamic preference for the deprotonated \ce{-SiO^-} and protonated \ce{-AlOH_2} state.

The second mechanism involves a solvent-assisted PT between two \ce{-SiOH} groups, where one site was already deprotonated to form \ce{-SiO^-} (see Supplementary Fig.~S10).
The corresponding 2D PTFEL reveals that the transition between stable states I and II lacks continuous low-energy pathways, consistent with an isolated, irreversible transfer event rather than dynamic back-and-forth exchange.
This interpretation is further supported by trajectory analysis, which shows that such transfers occurred only once or twice over the full 1.2~ns simulations, making it challenging to resolve a well-sampled free energy pathway for this mechanism.
Interestingly, a third local minimum (state~III) appears on the 2D landscape, corresponding to a stable but non-reactive solvation structure in which both hydrogen atoms of the bridging water molecule point toward surface groups.
This orientation is not conducive to PT, highlighting the influence of local hydrogen-bonding patterns.
These findings underscore that not all solvent-bridged geometries facilitate efficient PT: the directionality and strength of hydrogen bonds within the solvation network critically determine whether a bridging water molecule can support PT between surface groups.

Finally, while most solvent-assisted PT events involved a single bridging water molecule, we also observed double-water-mediated pathways under basic conditions. 
Representative snapshots in Fig.~\ref{fig:figure4}b illustrate such a multi-step transfer from \ce{-SiOH} to a neighboring \ce{-SiO^-}, mediated by two bridging water molecules. 
A similar process involving transfer from \ce{-AlOH_2} to \ce{-SiO^-} was observed in the same system (see Supplementary Fig.~S11).
Such double-bridge mechanisms have been reported previously PT between \ce{-AlOH_2} and \ce{-AlOH^-}~\cite{doi:10.1016/j.gca.2006.11.026} and between \ce{-SiOH} and \ce{-AlOH^-}~\cite{doi:10.1016/j.gca.2015.07.013}.
Our findings extend these mechanisms to a broader range of surface group combinations and edge contexts, demonstrating their general relevance at hydrated clay interfaces.
Notably, these events do not proceed via the transfer of a localized proton along a fixed structural pathway.
Instead, the transition occurs through a delocalized proton hole migrating across the hydrogen-bond network, which is characteristic of Grotthuss-like dynamics.
This behavior reveals a coupling between interfacial water dynamics, transient intermediate stabilization, and long-range proton transport at mineral–water interfaces.

\subsection{Isomorphic substitution impacts proton transfer energetics}
\begin{figure*}
    \centering{}
    \includegraphics[width=\textwidth]{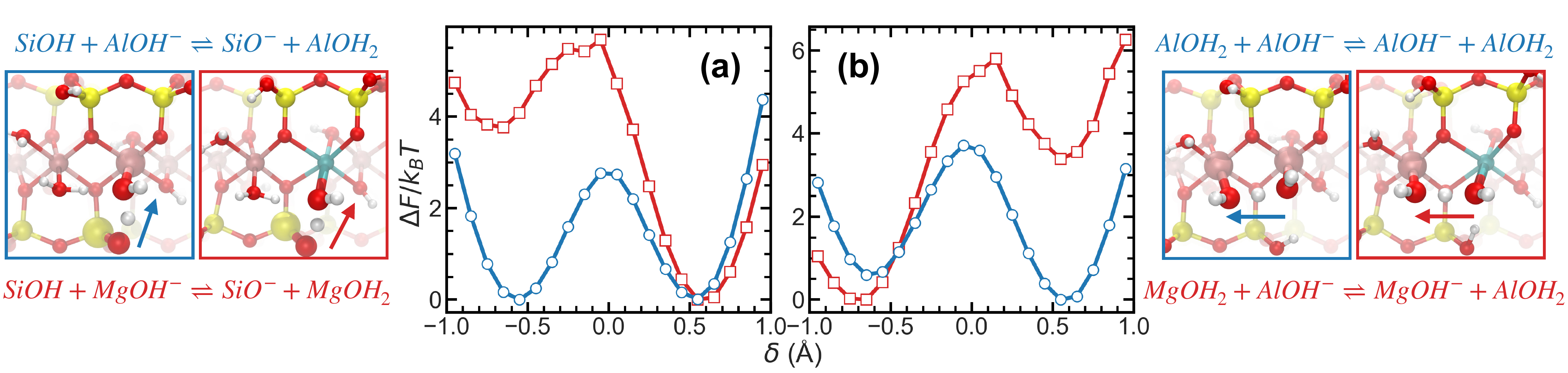}
    \caption{
        \textbf{Effect of isomorphic substitution (Mg for Al) on the free energy profiles of direct proton transfer reactions in the montmorillonite nanoparticle system.}
        Each panel compares two scenarios: red curves (highlighted by red frames) correspond to Mg-substituted sites, and blue curves (highlighted by blue frames) correspond to Al-only environments.
        For each case, the associated reaction equations and representative structures are shown alongside, with arrows indicating the direction of proton transfer.
        (a) Free energy profile for a direct proton transfer between \ce{-SiOH} and \ce{-AlOH^-}/\ce{-MgOH^-} within the site-2 region at the B/AC edge interface, under neutral conditions.
        (b) Free energy profile for a direct proton transfer between \ce{-AlOH_2}/\ce{-MgOH_2} and \ce{-AlOH^-} at the site-2 region under neutral conditions.
        }
    \label{fig:figure5}
\end{figure*}

Isomorphic substitution is an inherent feature of natural montmorillonite, where Mg$^{2+}$ ions replace Al$^{3+}$ in the octahedral layer.
In our study, three nanoparticle models were designed with distinct substitution patterns to systematically evaluate how Mg-for-Al substitution influences the reactivity of various edge surface sites (see Supplementary Section~S1).
These configurations allowed us to probe the energetic and mechanistic consequences of substitution at different crystallographic terminations. 

In acidic conditions, isomorphic substitution significantly enhances the reactivity of otherwise inert edge motifs.
As discussed above, the \ce{-AlAlOH} group, typically unreactive, upon substitution to form \ce{-AlMgOH} can be readily protonated by a hydronium ion from the bulk.
Notably, in neutral solution, the \ce{-AlMgOH} site was occasionally observed to transiently acquire a proton from a neighboring \ce{-SiOH} group, although this proton was often released shortly thereafter (see Supplementary Fig.~S13).
Similarly, both \ce{-SiAlO^-} and \ce{-SiMgO^-} groups are capable of undergoing protonation in acidic solution.
The corresponding PTFELs for reactions involving hydronium ions are shown in Supplementary Fig.~S12a.
For \ce{-SiAlO^-}, the free energy profile features two clear minima, corresponding to the protonated and deprotonated states.
In contrast, the profile for \ce{-SiMgO^-} exhibits only a single minimum, representing the protonated \ce{-SiMgOH} state.
This indicates a strong thermodynamic preference for proton retention at the Mg-substituted site.
The result is consistent with previously reported \textit{ab initio} pK$_a$ calculations, which place \ce{-SiMgOH} at 4.2 and \ce{-SiAlOH} at 1.7~\cite{doi:10.1016/j.gca.2014.05.044}.

Under basic conditions, isomorphic substitution similarly stabilizes protonated forms.
As shown in Supplementary Fig.~S12b, \ce{-MgOH_2} groups located on the AC edge (away from B-edge junctions) exhibit a pronounced asymmetry in their free energy profile: the barrier for deprotonation is substantially higher than that for reprotonation.
This thermodynamic bias implies that even after donating a proton to a hydroxide ion, \ce{-MgOH_2} quickly regains a proton from neighboring \ce{-SiOH}, maintaining its protonated state.
This observation is also consistent with previous first-principles molecular dynamics studies, which report that Mg-substituted hydroxyl groups on montmorillonite edges exhibit significantly elevated p$K_\mathrm{a}$ values (up to 13.2), indicating a strong preference for remaining protonated under typical pH conditions~\cite{doi:10.1016/j.gca.2013.04.008}.
Our simulations further reveal that despite this stability, \ce{-MgOH_2} groups can occasionally deprotonate and transiently participate in local proton transfer events under neutral conditions.
Such behavior, not captured by static p$K_\mathrm{a}$ values, highlights their potential role in dynamic interfacial charge redistribution.

In neutral solution, the effects of isomorphic substitution extend to the chain-like PT pathways at B/AC edge junctions.
Specifically, when the Al atom in the AC edge of a site~2 motif is replaced by Mg, the resulting \ce{-MgOH_2} group exhibits altered reactivity compared to its Al-based counterpart.
Fig.~\ref{fig:figure5}a presents the free energy profiles for direct PT between \ce{-SiOH} and \ce{-AlOH^-}/\ce{-MgOH^-}, while Fig.~\ref{fig:figure5}b shows PT between \ce{-AlOH_2}/\ce{-MgOH_2} and B-edge \ce{-AlOH^-}.
In both cases, Mg substitution increases the deprotonation barrier of the AC-edge donor group, thereby reducing its tendency to initiate PT.
Nonetheless, the two-state character of the free energy profiles is preserved, indicating that the overall PT sequence remains feasible: the B-edge \ce{-AlOH^-} acquires a proton from \ce{-MgOH_2}, which subsequently draws a proton from nearby \ce{-SiOH}, sustaining the direct chain-like transfer.
In this context, the Mg-substituted \ce{-MgOH_2} group acts as a transient proton relay, mediating net charge migration from \ce{-AlOH^-} to \ce{-SiO^-}.
However, the elevated deprotonation barrier substantially reduces the frequency and reversibility of such events, effectively suppressing surface PT dynamics in Mg-rich environments.
Taken together, these findings reveal that isomorphic substitution modulates PT dynamics in multiple ways: it promotes persistent protonation at specific sites while also altering local energy landscapes to suppress frequent exchange.
This dual role underscores the complex interplay between chemical composition and edge surface reactivity in natural clay minerals.

\section{Conclusions}
In this work, we employed MLP-MD simulations with first-principles accuracy to investigate the acid--base reactivity and PT dynamics of montmorillonite nanoparticles in aqueous environments over nanosecond timescales.
Our simulations reveal the mechanism of surface reactivity of clay edge surfaces and the widespread occurrence of both direct and solvent-assisted PT reactions among surface functional groups.
Through a systematic molecular-level analysis, we show that the amphoteric character of montmorillonite edge surfaces originates from the structural diversity of surface functional groups, which include both Brønsted acidic and basic moieties.
The observed dynamic proton transfer activity, especially under neutral conditions, arises from the intrinsic instability of \ce{-AlOH^-} groups on the (010) B edge, which readily engage in proton exchange processes.
We further identify that basic conditions promote PT activity, while isomorphic substitution of Al by Mg increases deprotonation barriers and thereby stabilizes protonated surface states.
These findings provide a molecular-level complement to macroscopic titration experiments and surface complexation models, which cannot unambiguously resolve site-specific reactivity or dynamic proton exchange.
Our results show that clay edges are not static arrays of surface groups with fixed p$K_\mathrm{a}$ values, but dynamic, proton-conducting networks whose chemistry is tunable by both composition and environment.
By explicitly simulating the interactions between edge functional groups and aqueous acid--base species, our work offers the first direct mapping of the reactive sites, mechanistic pathways, and characteristic timescales governing surface charge regulation in montmorillonite.
This capability enables a more complete and predictive understanding of clay surface chemistry under realistic environmental conditions.

Our results move beyond previous static or short-timescale simulations by capturing the dynamic and recurrent nature of surface PT reactions, including previously 
unreported
inter-site mechanisms.
By identifying the influence of both surface topology and isomorphic substitution on reactivity, our study clarifies why certain sites remain inert under acidic conditions while others remain persistently active under basic conditions.
The observed solvent-assisted mechanisms further highlight the importance of the surrounding hydrogen-bonding network in modulating PT barriers and enabling transient charge delocalization along clay edges.
These insights support a more realistic picture of clay edge reactivity, where protons are not localized but can hop across a network of labile sites.
This proton mobility is expected to play a key role in charge buffering, interfacial conductivity, and dynamic adsorption behavior in natural and engineered clay-based systems.

The simulation framework developed here can be readily extended to larger-scale clay interfaces and composite systems, enabling long-timescale tracking of surface reactivity, dissolution, and nanoparticle aggregation phenomena.
A deeper understanding of surface proton dynamics will facilitate the development of predictive models for the adsorption and desorption of metal cations, organics, and pollutants on clay minerals~\cite{doi:10.1021/acs.est.6b04677,doi:10.1007/s11356-020-10985-9}.
Moreover, the dynamic nature of hydrogen-bond networks and mobile protons at clay edges may provide design insights for catalytic applications involving activated clays~\cite{doi:10.1016/j.clay.2015.12.005,doi:10.3390/min14060629}, as well as for optimizing the ionic conductivity and surface charge balance of clay-based materials in energy storage devices~\cite{doi:10.1002/advs.202004036}.
These findings underscore the central role of edge-resolved, proton-mediated mechanisms in shaping the broader physicochemical behaviors of clay mineral systems.

\section{Methods}

\subsection{Machine Learning Potentials}
The machine learning potential used in this work was developed within the MACE framework~\cite{doi:10.48550/arXiv.2206.07697}, which combines message passing with high body-order equivariant features to accurately capture many-body interactions.
This architecture has demonstrated excellent transferability and accuracy across diverse chemical systems~\cite{doi:10.1063/5.0155322}.
For this study, the MACE model was configured with two message-passing layers and four-body equivariant features, using a radial cutoff of 5~\AA.
Although explicit long-range interactions are not directly included, the receptive field of the model—defined by the product of the cutoff and the number of layers—effectively reaches 10~\AA, which is sufficient to capture the relevant short- and medium-range interactions at clay–water interfaces.

The training dataset was generated from DFT calculations using the CP2K simulation package~\cite{doi:10.1063/5.0007045}, based on the Gaussian and plane wave (GPW) method with a high plane-wave cutoff of 1200~Ry~\cite{doi:10.1021/acs.jpclett.4c01030}.
The revPBE functional~\cite{doi:10.1103/PhysRevLett.77.3865, doi:10.1103/PhysRevLett.80.890} combined with Grimme’s D3 dispersion correction~\cite{doi:10.1063/1.3382344} was adopted to provide a balanced description of hydrogen-bonded liquids and silicate solids~\cite{doi:10.1021/acs.jpcc.6b09559}.
This choice has been validated in prior studies for modeling both the structure and dynamics of water~\cite{doi:10.1021/acs.jpclett.7b00391} and the vibrational and mechanical properties of clay minerals~\cite{doi:10.1063/5.0152361}.
Electron–ion interactions were treated using Goedecker–Teter–Hutter (GTH) pseudopotentials~\cite{doi:10.1103/PhysRevB.54.1703} with element-specific basis sets: TZV2P-GTH for H, O, Na, and Cl, and DZVP-MOLOPT-SR-GTH for Al, Mg, and Si.
All calculations employed periodic boundary conditions, and simulation cells were constructed to match the structural features of each system, while retaining charge neutrality.

The training dataset encompassed bulk aqueous solutions, diverse montmorillonite surfaces, and interface structures under different pH conditions, capturing both structural and chemical variability (see Supplementary Section~S1).
The trained model achieves low root mean squared errors (RMSEs) of 0.4~meV/atom in energies and 26.2~meV/\AA{} in forces on the training set, with similarly low values on the test set (0.5~meV/atom and 27.7~meV/\AA).
Further validation confirmed the model’s reliability in reproducing interfacial energetics, clay lattice structures, and bulk water properties (see Supplementary Section~S2).

\subsection{Molecular Dynamics Simulations}
MLP-based molecular dynamics simulations were conducted in LAMMPS\cite{doi:10.1016/j.cpc.2021.108171} under NPT conditions at 298~K and 1.01325~bar using the Nosé–Hoover thermostat and barostat.
A time step of 0.5~fs was used, and three central oxygen atoms of each clay particle were constrained to preserve structural integrity.
Simulations were performed on three montmorillonite nanoparticles, each embedded in acidic, neutral, and basic aqueous environments, resulting in nine total systems.
Each trajectory was run for 1.2~ns with periodic boundary conditions in all directions, and representative configurations were extracted for further analysis,
resulting in total in 10\,ns for analysis.

\subsection{Proton Transfer Free Energy Landscape}
PT events at montmorillonite edges were analyzed by constructing one- and two-dimensional free energy profiles from molecular dynamics trajectories.
Two types of PT mechanisms were considered: direct PT between neighboring surface sites and solvent-assisted transfer mediated by bridging water molecules.

For direct PT, hydrogen bonds were first identified based on geometric criteria (O$_\mathrm{d}$--O$_\mathrm{a} <$ 3.5~\AA{}, and O$_\mathrm{a}$--O$_\mathrm{d}$--H$_\mathrm{d} <$ 30$^\circ$)\cite{doi:10.1063/1.2431168}.
A reaction coordinate $\delta$ was defined for each hydrogen-bond donor as:
\begin{equation}
    \delta = \min(d_{\mathrm{Hd}-\mathrm{Oa}} - d_{\mathrm{Od}-\mathrm{Hd}})
\end{equation}
and used to construct free energy profiles via:
\begin{equation}
\Delta F(\delta)/k_B T = - \ln P(\delta)
\end{equation}
where $P(\delta)$ is the normalized probability distribution.
Signs of $\delta$ were assigned to distinguish forward and reverse reactions, as annotated in each figure.

For solvent-assisted PT, configurations involving a water molecule bridging two reactive edge sites (O$_1$ and O$_3$) were identified, with O$_2$ denoting the central bridging oxygen.
Two reaction coordinates were defined to describe the proton positions relative to their donor and acceptor oxygen atoms:
\begin{equation}
\xi_1 = d_{\mathrm{O2}-\mathrm{H1}} - d_{\mathrm{O1}-\mathrm{H1}}, \quad 
\xi_2 = d_{\mathrm{O2}-\mathrm{H2}} - d_{\mathrm{O3}-\mathrm{H2}}
\end{equation}

For one-dimensional analysis, we used the averaged coordinate $\xi = (\xi_1 + \xi_2)/2$.
For two-dimensional profiles, the joint distribution $P(\xi_1, \xi_2)$ was used to compute:
\begin{equation}
    \Delta F(\xi_1, \xi_2)/k_B T = - \ln P(\xi_1, \xi_2)
\end{equation}

Complete definitions, identification criteria, and sampling procedures are detailed in Supplementary Section~S4.

\begin{acknowledgments}
We would like to thank Professor Guoping Zhang for early discussions that helped shape the initial ideas of this study.  
Y.F. acknowledges support from the National Key Research and Development Program of China (2022YFC3201803) and China Scholarship Council.
X.R.A. acknowledges support from the European Union under the "n-AQUA" European Research Council project (Grant No. 101071937).
C.S. acknowledges financial support from the Royal Society, grant number RGS/R2/242614.
Calculations were performed on the Cambridge Service for Data Driven Discovery (CSD3) operated by the University of Cambridge Research Computing Service (www.csd3.cam.ac.uk) and the Center for Computational Science and Engineering at the Southern University of Science and Technology.
%
\end{acknowledgments}

\section*{Competing interests}
The authors declare no competing interests.

\section*{References}

\bibliographystyle{aipnum4-2}
\bibliography{refs}


\end{document}


\maketitle
\begin{center}
    \textit{$^{a}$~State Key Laboratory of Hydro-science and Engineering, Department of Hydraulic Engineering, Tsinghua University, Beijing 100084, China.\\}
    \textit{$^{b}$~Yusuf Hamied Department of Chemistry, University of Cambridge, Lensfield Road, Cambridge, CB2 1EW, UK.\\}
    \textit{$^{c}$~Cavendish Laboratory, Department of Physics, University of Cambridge, Cambridge, CB3 0HE, UK.\\}
    \textit{$^{d}$~Lennard-Jones Centre, University of Cambridge, Trinity Ln, Cambridge, CB2 1TN, UK.\\}
    \textit{E-mail: fanghw@tsinghua.edu.cn, cs2121@cam.ac.uk}
\end{center}
\newpage
\onehalfspacing
\tableofcontents
\newpage
\section{Development of Machine Learning Potential}
\subsection{Montmorillonite Nanoparticles}
The montmorillonite structures used in this study were constructed based on experimental data~\cite{doi:10.1524/zkri.1971.134.16.196}, incorporating isomorphic substitutions of magnesium for aluminum within the octahedral sheet. The nanoparticles were constructed to predominantly expose the (110) and (010) edge surfaces, herein referred to as the AC and B edges, respectively~\cite{doi:10.1346/CCMN.1988.0360207}. For clarity, these designations (AC and B) are consistently used throughout this work. The AC and B edge surfaces constitute approximately 60\% and 20\% of the total edge area~\cite{doi:10.1016/j.clay.2020.105442}, respectively, and are considered among the most stable terminations for montmorillonite~\cite{doi:10.1346/CCMN.2015.0630403, doi:10.3390/min6020025, doi:10.1016/j.gca.2016.06.021}.

To comprehensively investigate the interfacial properties between montmorillonite and aqueous solution, three distinct nanoparticle structures (denoted as Mont.1, Mont.2, and Mont.3) were developed. These structures share an identical total number of isomorphic substitutions (three Mg-for-Al substitutions), but differ in the spatial distribution of the substitution sites, which were assigned randomly. The isomorphic substitutions located within the particle interior and on the AC edges were charge-balanced by Na\textsuperscript{+} counterions. In contrast, substitutions at the B edges were compensated by protonation of Mg atoms, forming terminal –(OH\textsubscript{2})\textsubscript{2} groups. For Al atoms situated at these edge sites, one –OH\textsubscript{2} and one –OH group were formed. The detailed structural formulas and compositional information of each nanoparticle model are summarized in Table~\ref{tab:mont_structures}.

\begin{table}[!htbp]
    \centering
    \caption{Compositional and structural characteristics of the montmorillonite nanoparticles.}
    \label{tab:mont_structures}
    \begin{tabular}{
        >{\raggedright\arraybackslash}m{4.8cm} |
        >{\raggedright\arraybackslash}m{4.2cm} |
        >{\centering\arraybackslash}m{4.5cm}
    }
        \hline
        \textbf{Structure} & \textbf{Isomorphic Substitution Pattern} & \textbf{Visualization} \\
        \hline
        Mont.1 \newline Na$_2$HSi$_{48}$(Al$_{21}$Mg$_3$)O$_{96}$(OH)$_{72}$ & 
        $N_{\text{inner}} = 2$ \newline $N_{B\ \text{edge}} = 1$ & 
        \includegraphics[height=3cm]{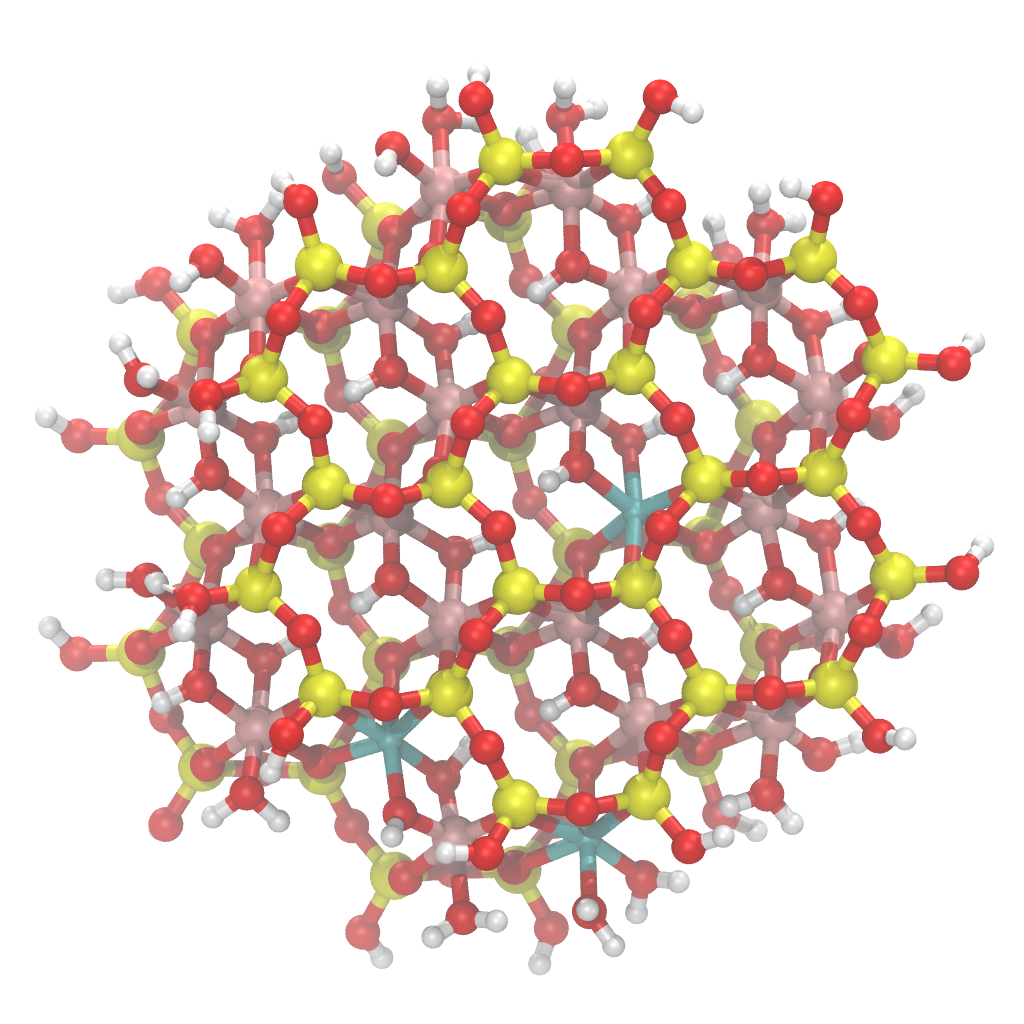} \\
        \hline
        Mont.2 \newline Na$_3$Si$_{48}$(Al$_{21}$Mg$_3$)O$_{96}$(OH)$_{72}$ & 
        $N_{\text{inner}} = 1$ \newline $N_{AC\ \text{edge}} = 2$ & 
        \includegraphics[height=3cm]{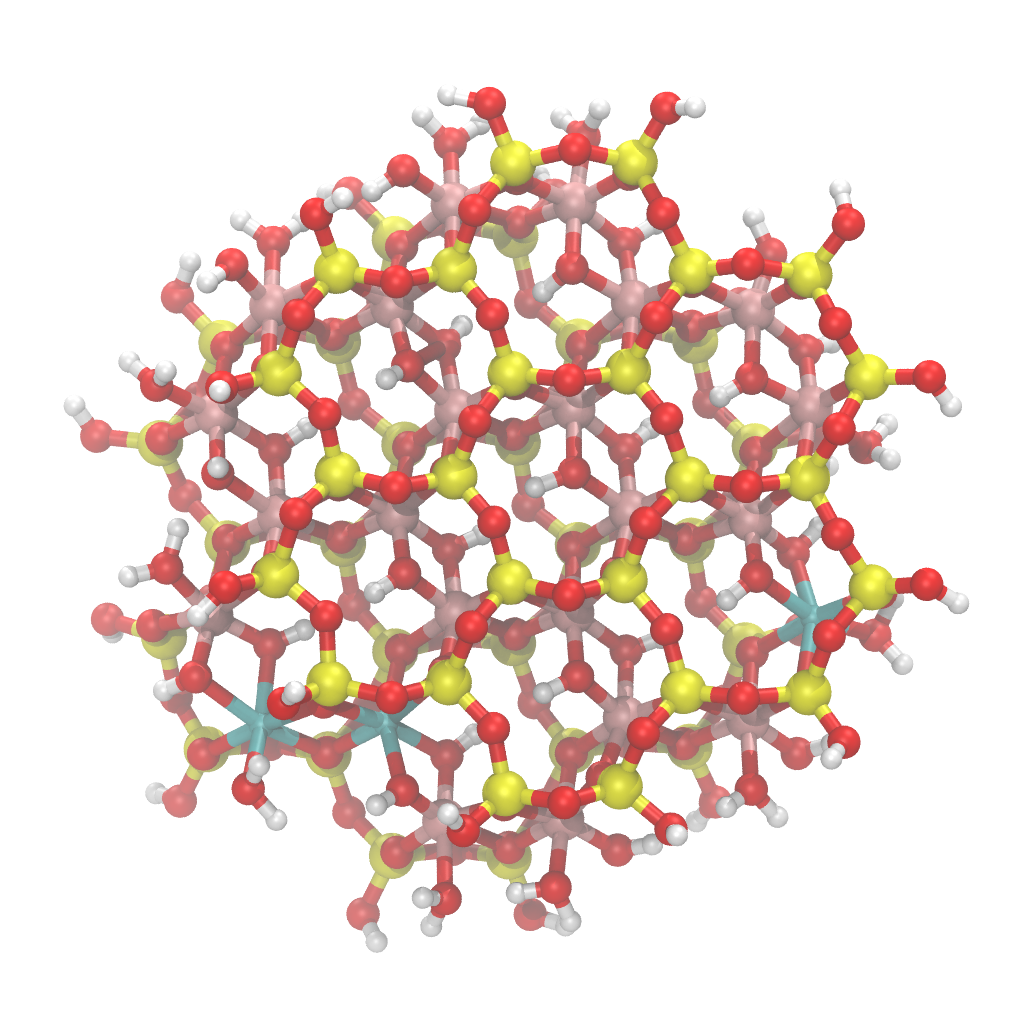} \\
        \hline
        Mont.3 \newline Na$_2$HSi$_{48}$(Al$_{21}$Mg$_3$)O$_{96}$(OH)$_{72}$ & 
        $N_{\text{inner}} = 1$ \newline $N_{AC\ \text{edge}} = 1$ \newline $N_{B\ \text{edge}} = 1$ & 
        \includegraphics[height=3cm]{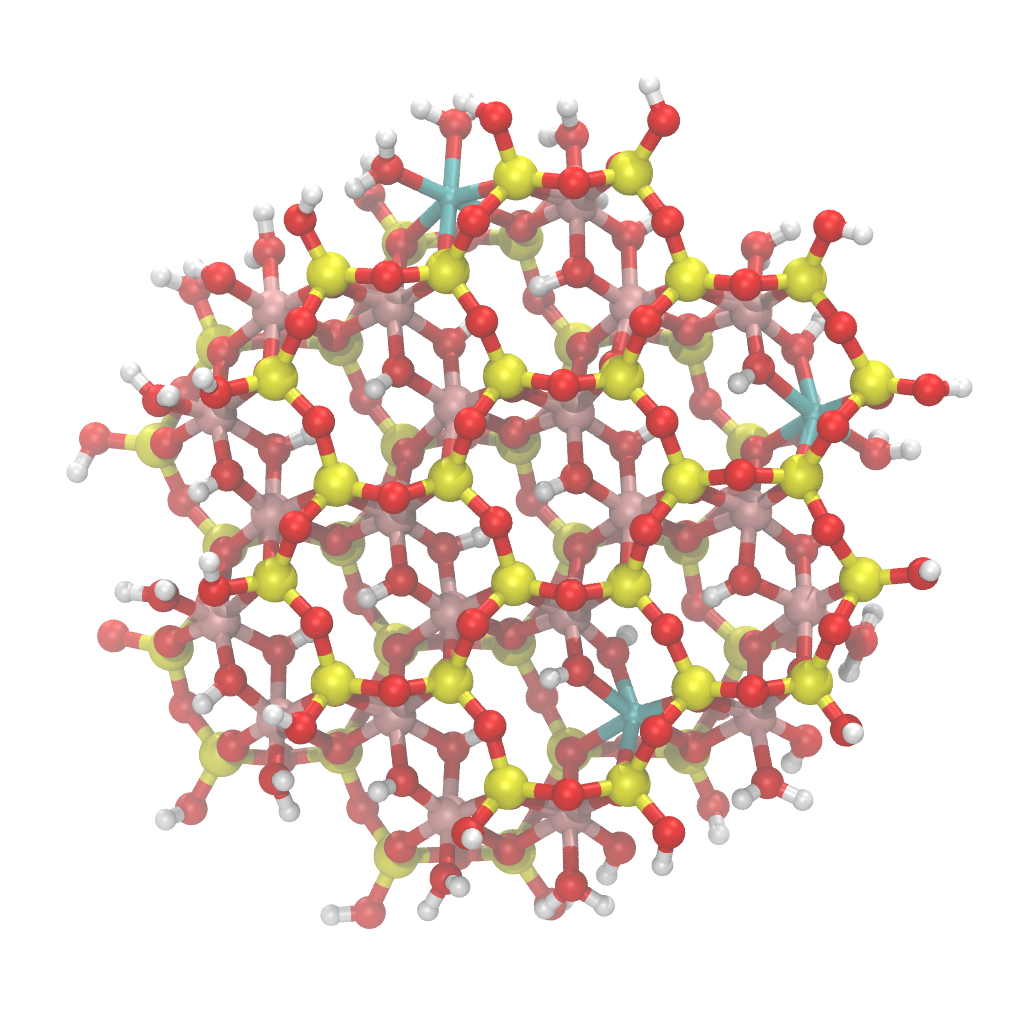} \\
        \hline
    \end{tabular}
\end{table}

\subsection{Dataset Construction}
To develop a machine learning potential (MLP) capable of accurately modeling the complex montmorillonite–aqueous solution interface, the dataset must comprehensively represent a wide range of relevant chemical environments. These include bulk aqueous solutions and explicit solid–liquid interface systems. Furthermore, to ensure that the resulting MLP has sufficient generalization ability and transferability to larger-scale interfaces and various montmorillonite structures, the dataset was designed to include diverse system sizes, aqueous solutions with different pH conditions, and montmorillonite nanoparticles with different distributions of isomorphic substitutions. A detailed summary of the dataset composition is provided in Table~\ref{tab:dataset_config}.

All configurations were generated through molecular dynamics (MD) simulations performed in the NPT ensemble using the LAMMPS package\cite{doi:10.1016/j.cpc.2021.108171} and starting from the MACE-MP-0 foundation model~\cite{doi:10.48550/arXiv.2401.00096}.
We also generated further training configurations in a reinforcement cycle using preliminary MACE models trained to the target revPBE-D3 reference to improve data set coverage.
The simulations were conducted at temperatures ranging from 298~K to 400~K with a time step of 0.5~fs. Temperature and pressure were controlled using Nosé–Hoover thermostat and barostat with damping parameters of 50~fs and 500~fs. For each system, 30–40 structures were randomly selected from the MD trajectories, resulting in a dataset consisting of 580 unique atomic configurations. Although the number of unique configurations is moderate, the resulting dataset includes a total of 2,417,016 force components, providing ample coverage for model training and validation.

\begin{longtable}{>{\raggedright\arraybackslash}m{4cm}|>{\raggedright\arraybackslash}m{3cm}|>{\raggedright\arraybackslash}m{3cm}|>{\centering\arraybackslash}m{4cm}}
    \caption{Details of the dataset} \\
    \hline
    \label{tab:dataset_config}
    System & Number & Simulation details & Visualization \\ \hline
    \endfirsthead
    \hline
    System & Number & Simulation details & Visualization \\ \hline
    \endhead
    \hline
    \endfoot
    \hline
    \endlastfoot
    Bulk water & 40 & N$_{water}$ = 300  & \includegraphics[height=2.8cm]{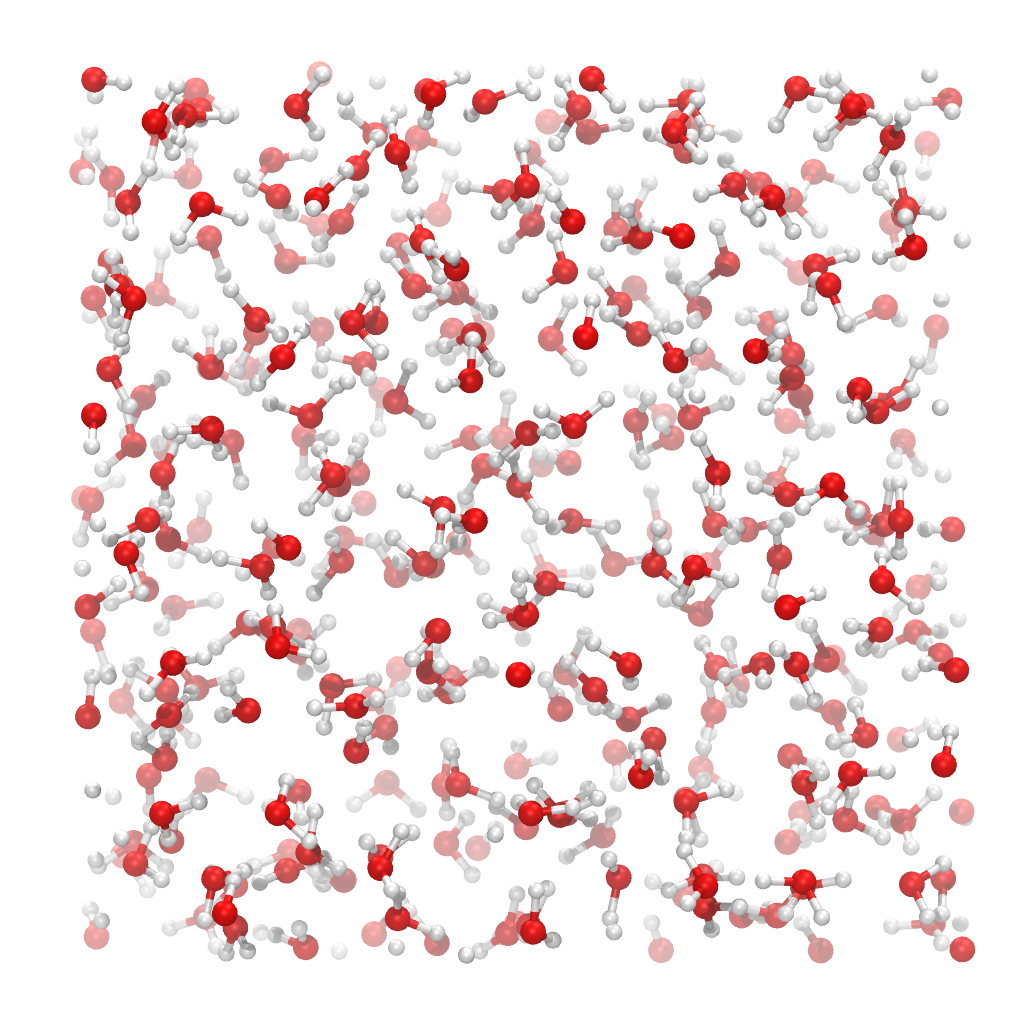} \\ \hline
    NaCl solution & 30 & N$_{water}$ = 500 \newline N$_{Na^+}$ = 10 \newline N$_{Cl^-}$ = 10 & \includegraphics[height=2.8cm]{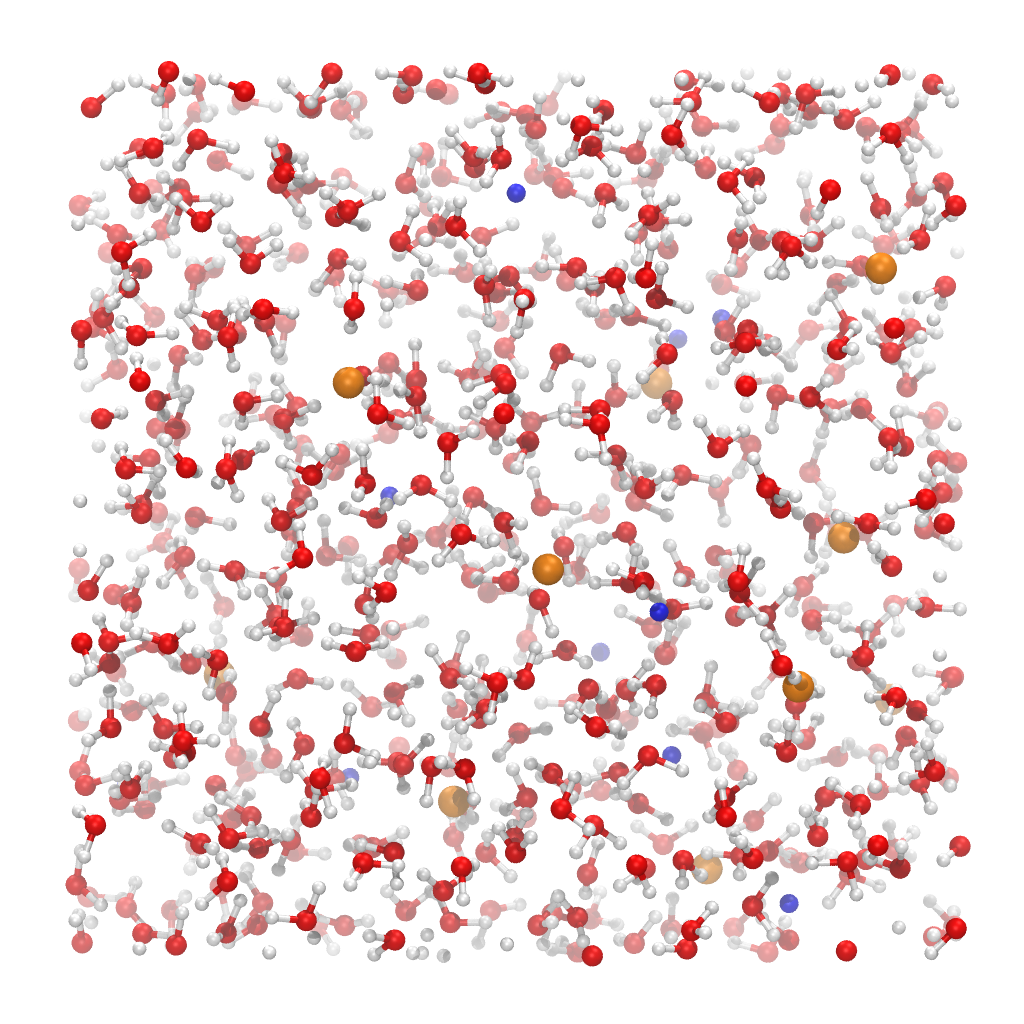} \\ \hline
    NaCl-HCl solution & 30 & N$_{water}$ = 500 \newline N$_{Na^+}$ = 5 \newline N$_{H3O^+}$ = 5 \newline N$_{Cl^-}$ = 10 & \includegraphics[height=2.8cm]{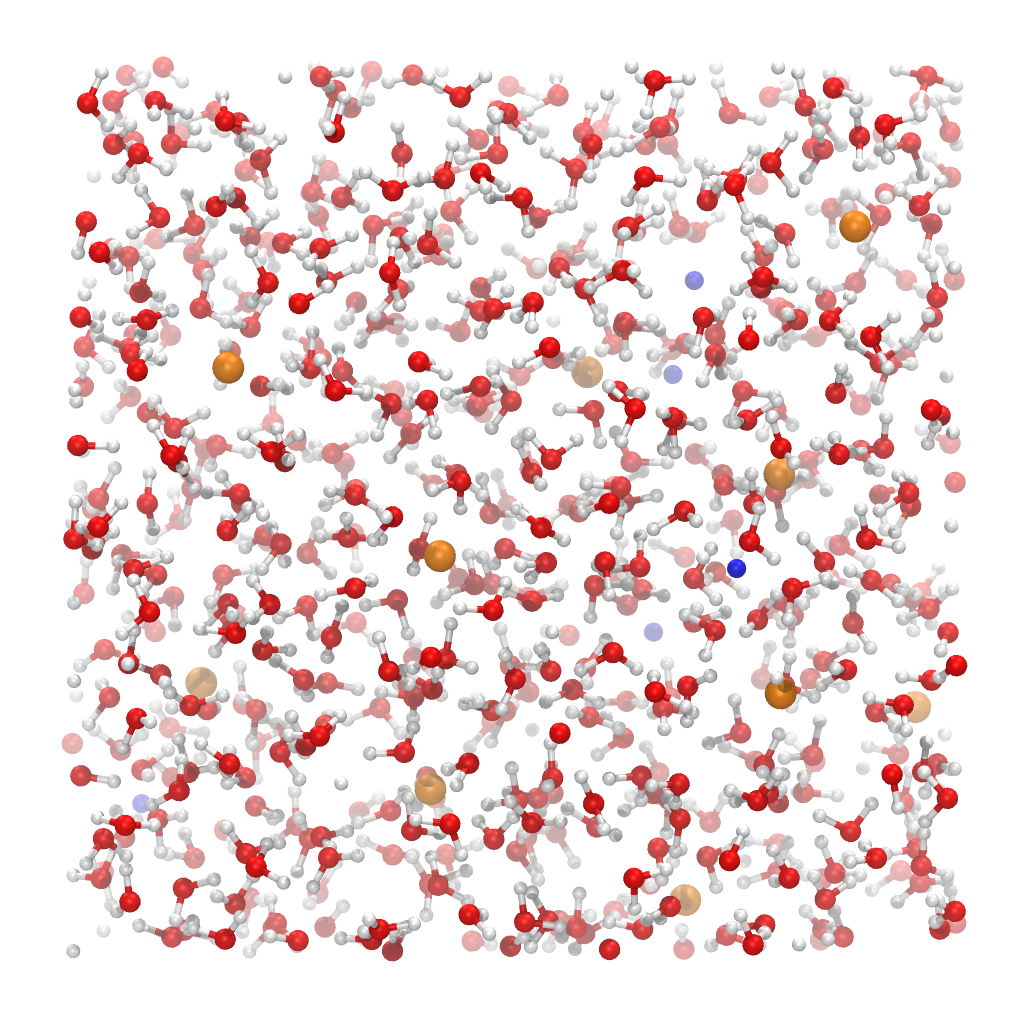} \\ \hline
    NaCl-NaOH solution & 30 & N$_{water}$ = 500 \newline N$_{Na^+}$ = 10 \newline N$_{OH^-}$ = 5 \newline N$_{Cl^-}$ = 5 & \includegraphics[height=2.8cm]{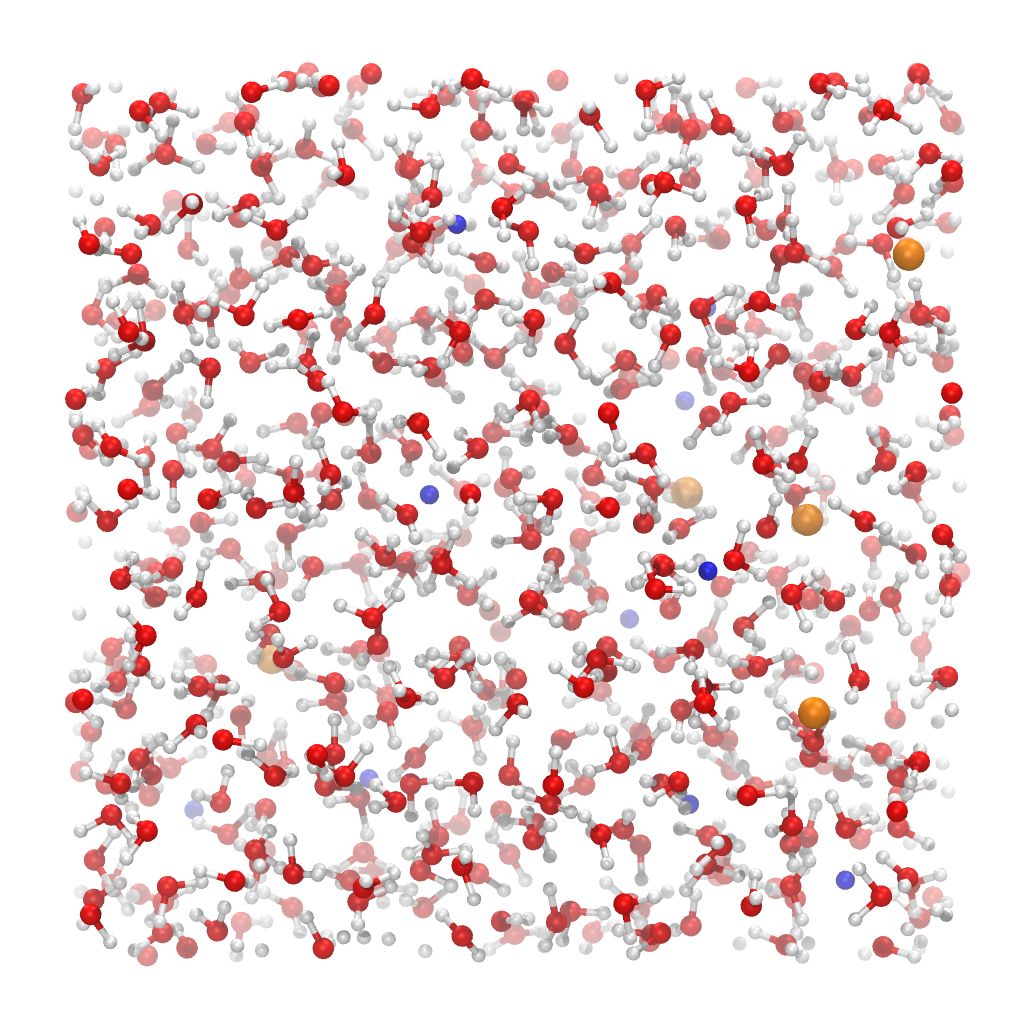} \\ \hline
    NaCl pair & 30 & N$_{water}$ = 300 \newline N$_{Na^+}$ = 1 \newline N$_{Cl^-}$ = 1 & \includegraphics[height=2.8cm]{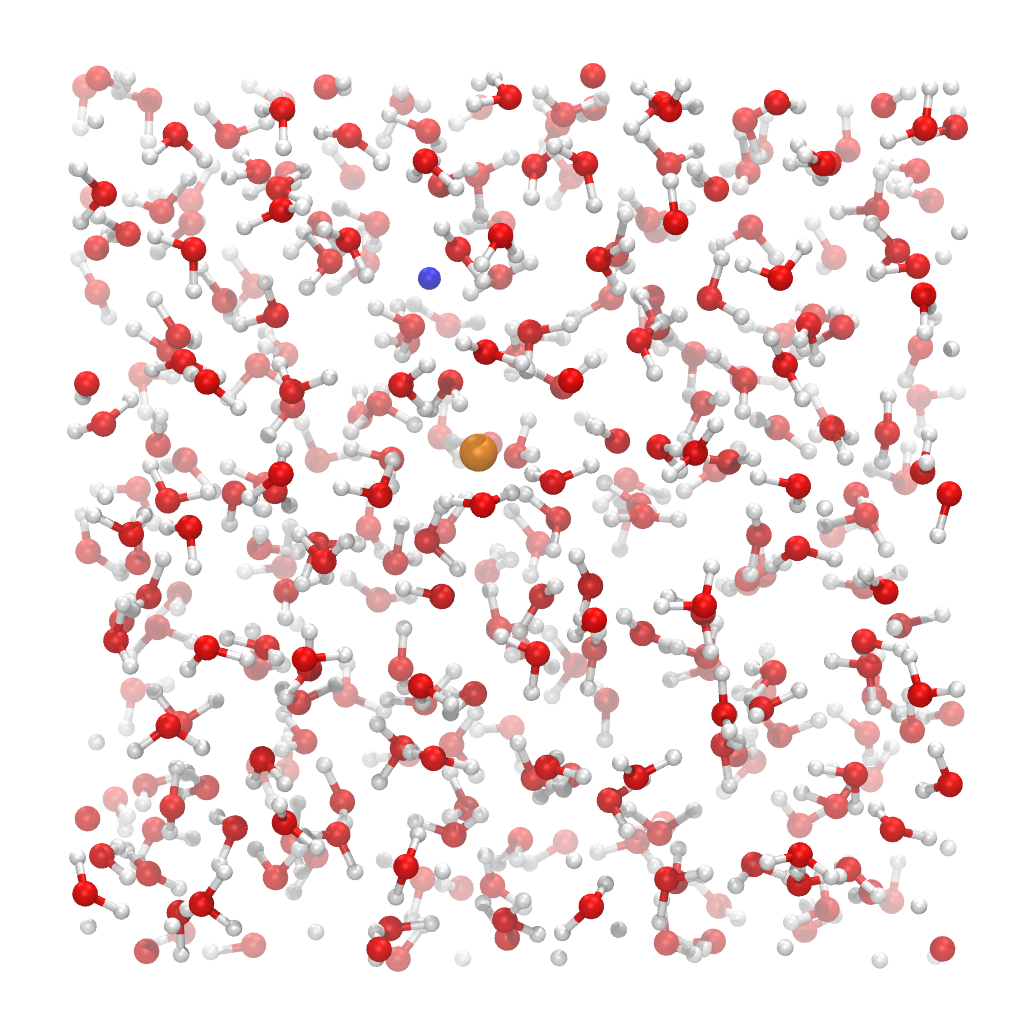} \\ \hline
    HCl pair & 30 & N$_{water}$ = 300 \newline N$_{H3O^+}$ = 1 \newline N$_{Cl^-}$ = 1 & \includegraphics[height=2.8cm]{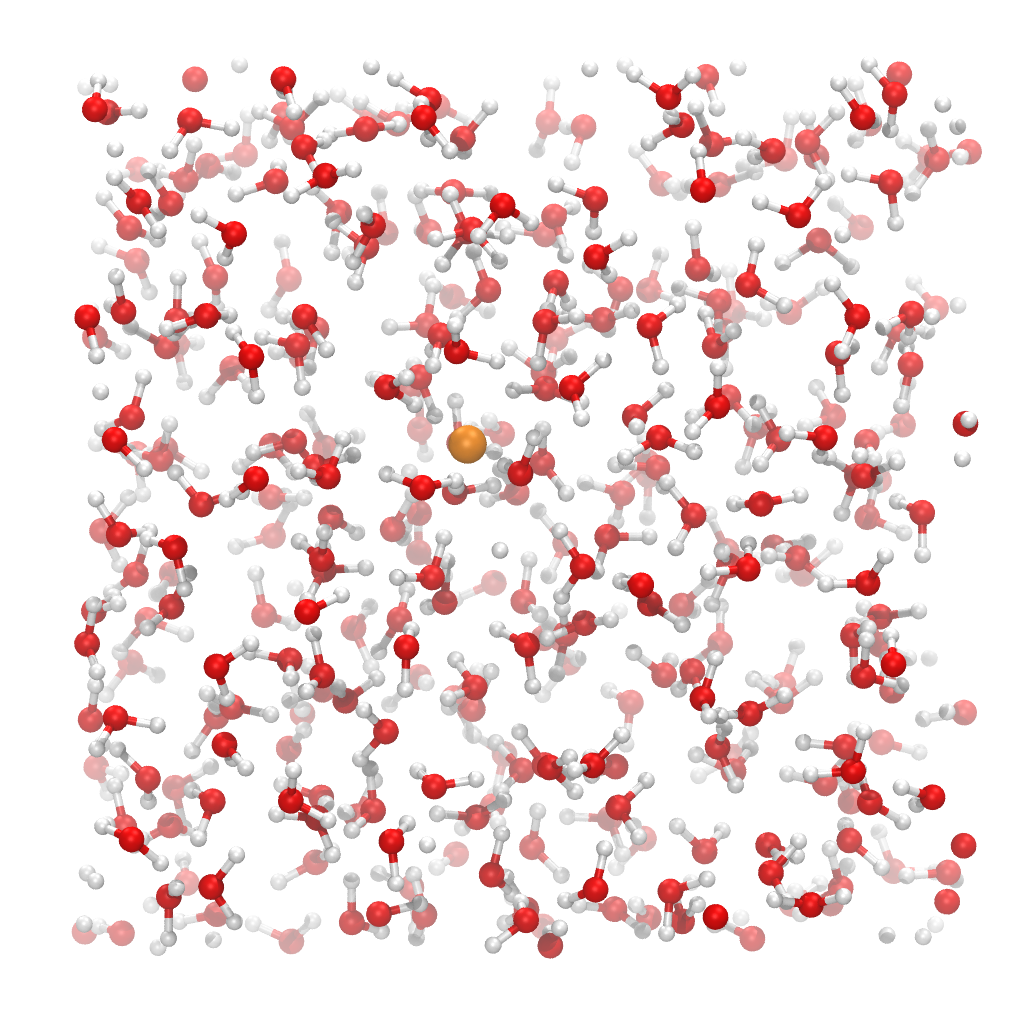} \\ \hline
    NaOH pair & 30 & N$_{water}$ = 300 \newline N$_{Na^+}$ = 1 \newline N$_{OH^-}$ = 1 & \includegraphics[height=2.8cm]{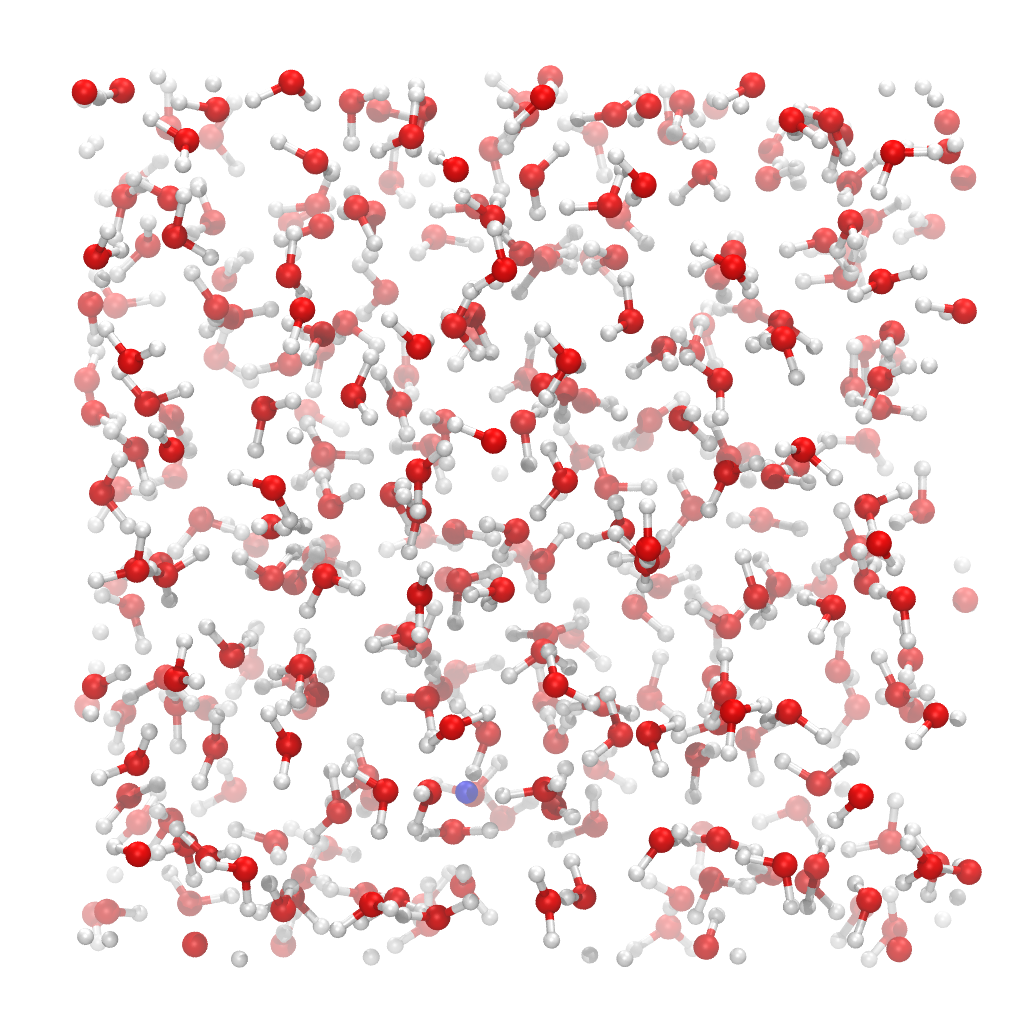} \\ \hline
    Si$_{48}$Al$_{24}$O$_{96}$(OH)$_{72}$–water interface & 40 & N$_{Mont.\ particle}$ = 1 \newline N$_{water}$ = 250 & \includegraphics[height=2.8cm]{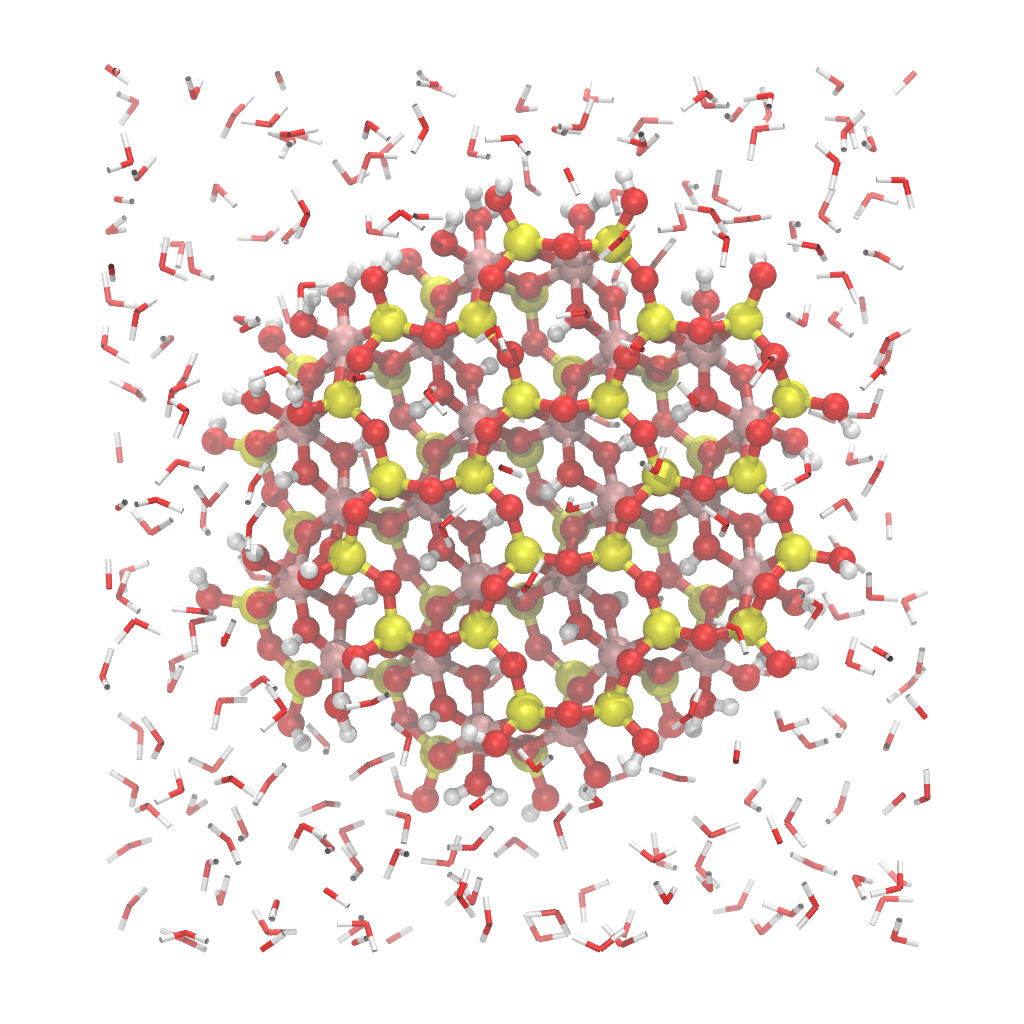} \\ \hline
    Si$_{48}$Al$_{24}$O$_{96}$(OH)$_{72}$–acid solution interface & 40 & N$_{Mont.\ particle}$ = 1 \newline N$_{water}$ = 250 \newline N$_{H3O^+}$ = 3 \newline N$_{Cl^-}$ = 3 & \includegraphics[height=2.8cm]{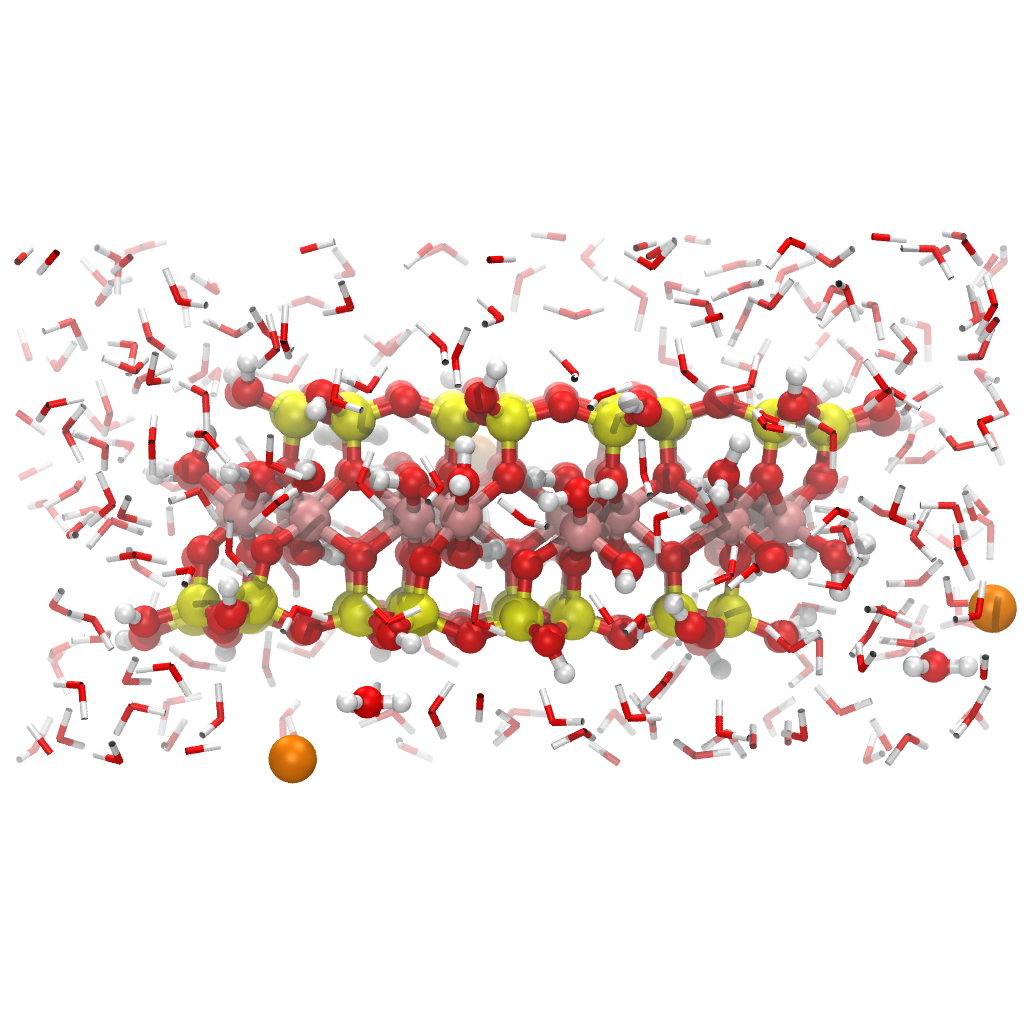} \\ \hline
    Si$_{48}$Al$_{24}$O$_{96}$(OH)$_{72}$–base solution interface & 40 & N$_{Mont.\ particle}$ = 1 \newline N$_{water}$ = 250 \newline N$_{Na^+}$ = 3 \newline N$_{OH^-}$ = 3 & \includegraphics[height=2.8cm]{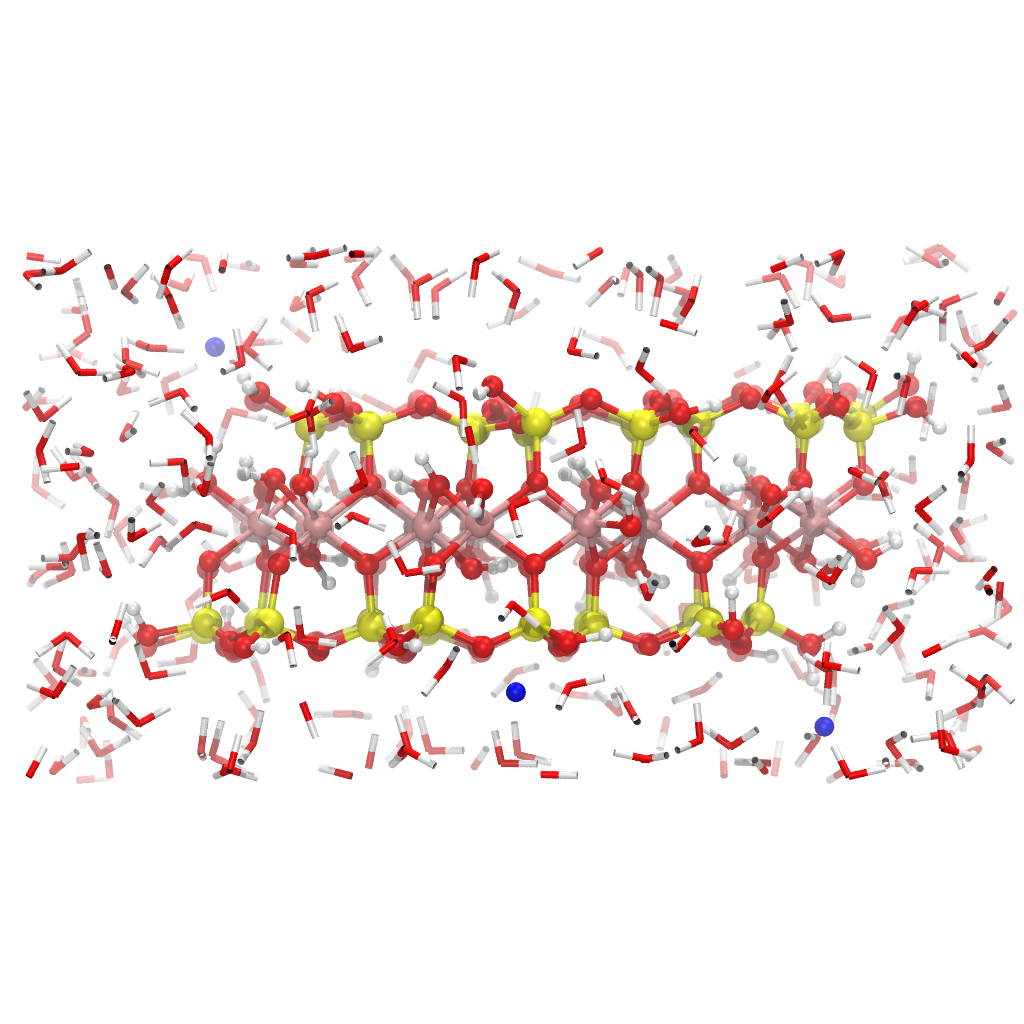} \\ \hline
    Mont.1–water interface & 40 & N$_{Mont.\  particle}$ = 1 \newline N$_{water}$ = 250 & \includegraphics[height=2.8cm]{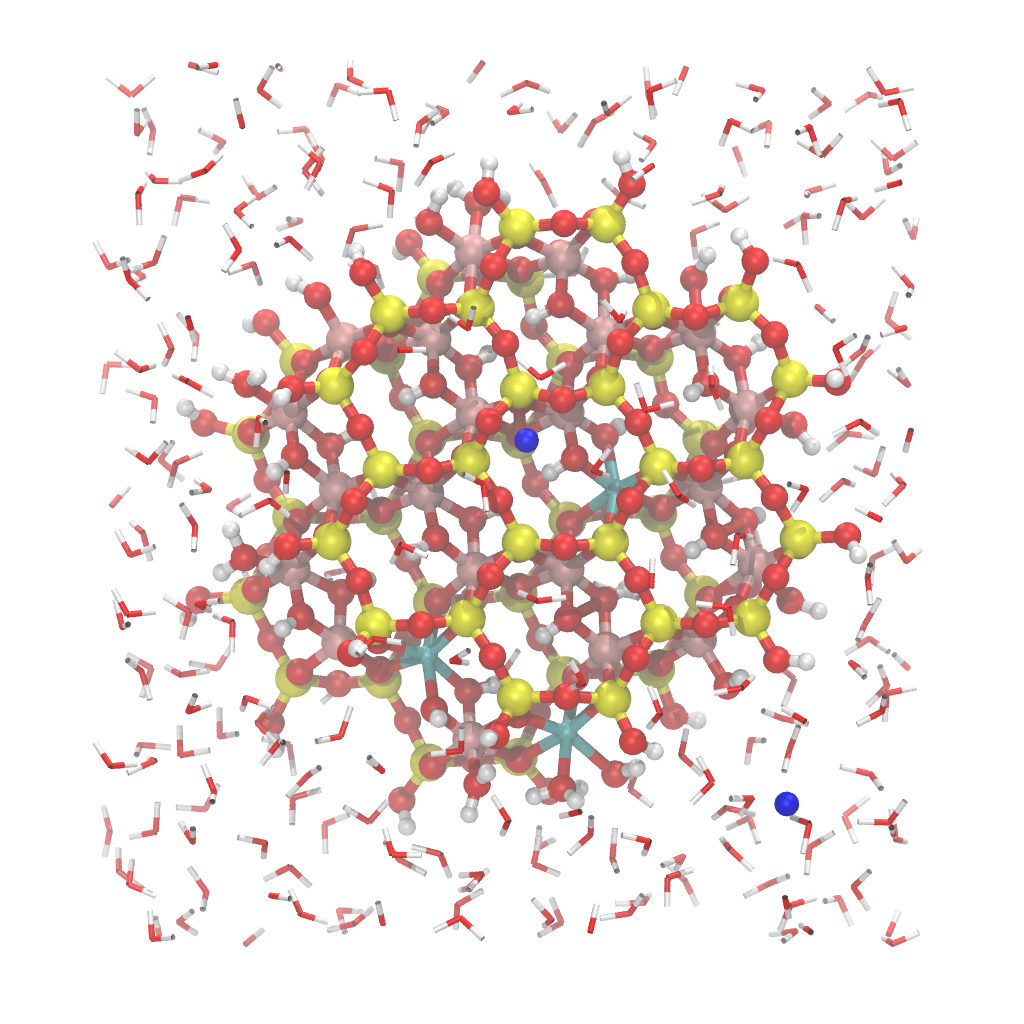} \\ \hline
    Mont.1–acid solution interface & 40 & N$_{Mont.\ particle}$ = 1 \newline N$_{water}$ = 250 \newline N$_{H3O^+}$ = 3 \newline N$_{Cl^-}$ = 3 & \includegraphics[height=2.8cm]{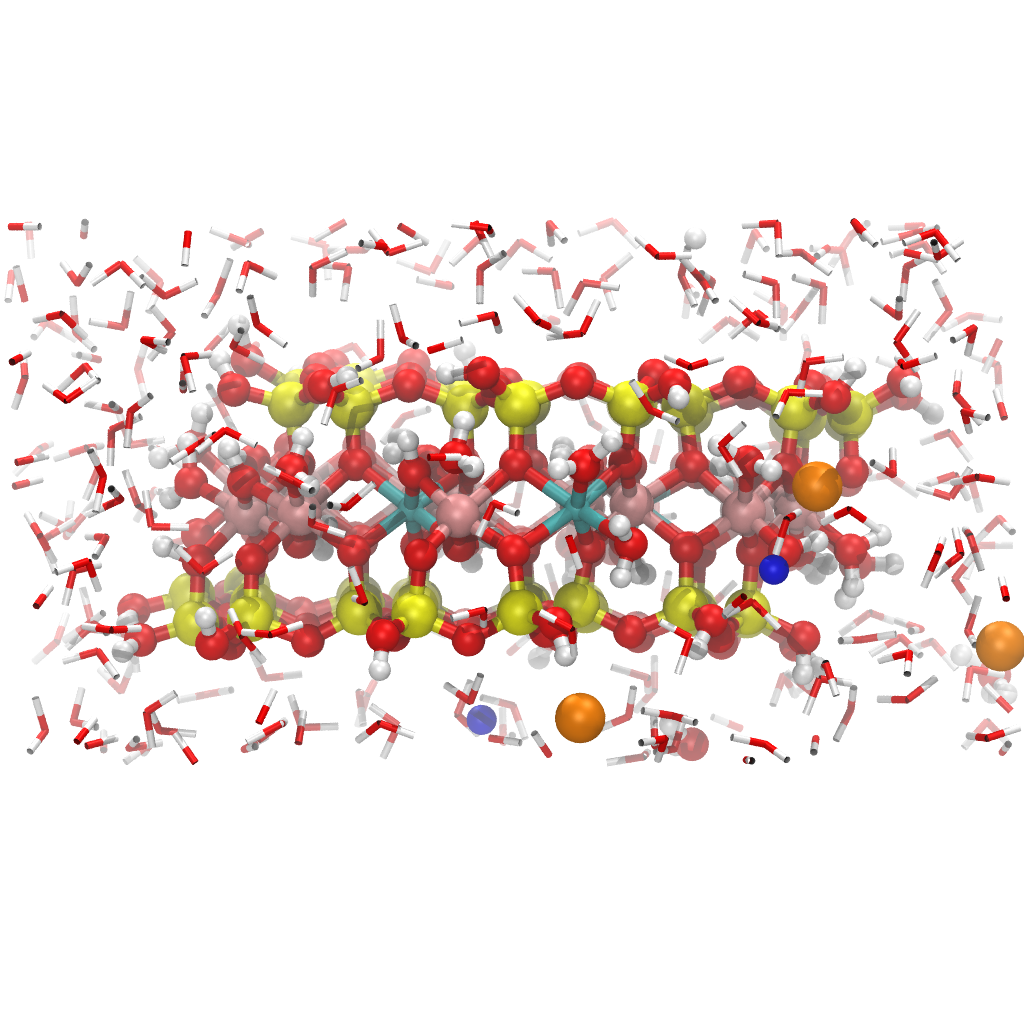} \\ \hline
    Mont.1–base solution interface & 40 & N$_{Mont.\ particle}$ = 1 \newline N$_{water}$ = 250 \newline N$_{Na^+}$ = 3 \newline N$_{OH^-}$ = 3 & \includegraphics[height=2.8cm]{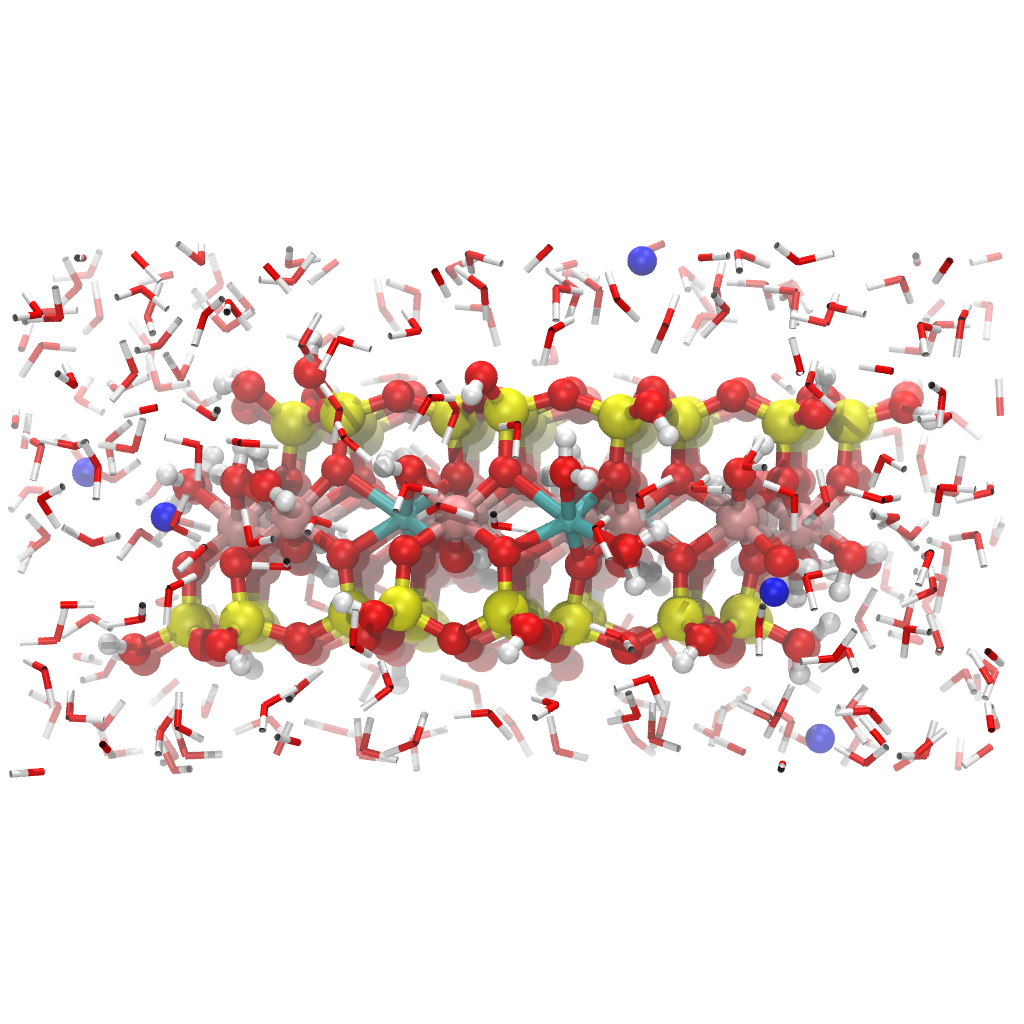} \\ \hline
    Mont.2–water interface & 40 & N$_{Mont.\ particle}$ = 1 \newline N$_{water}$ = 750 & \includegraphics[height=2.8cm]{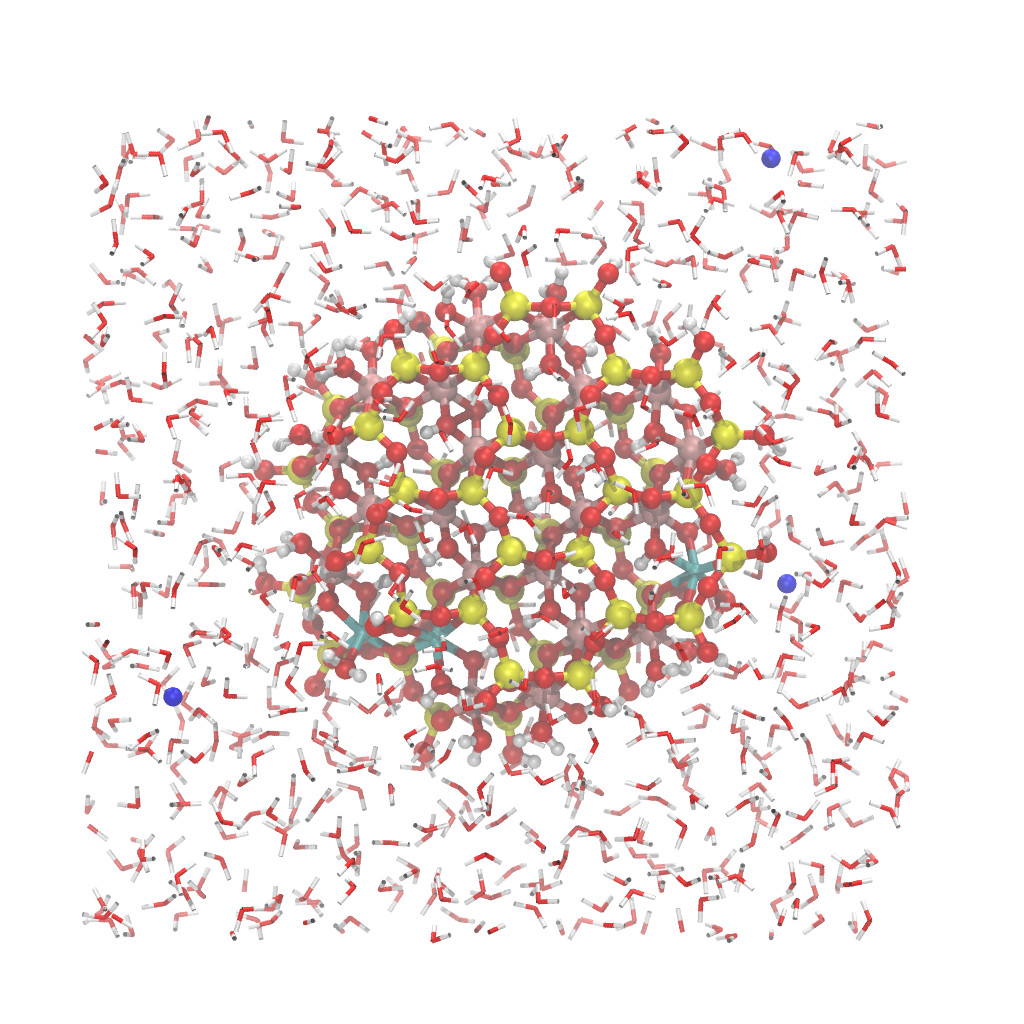} \\ \hline
    Mont.2–acid solution interface & 40 & N$_{Mont.\ particle}$ = 1 \newline N$_{water}$ = 750 \newline N$_{H3O^+}$ = 5 \newline N$_{Cl^-}$ = 5 & \includegraphics[height=2.8cm]{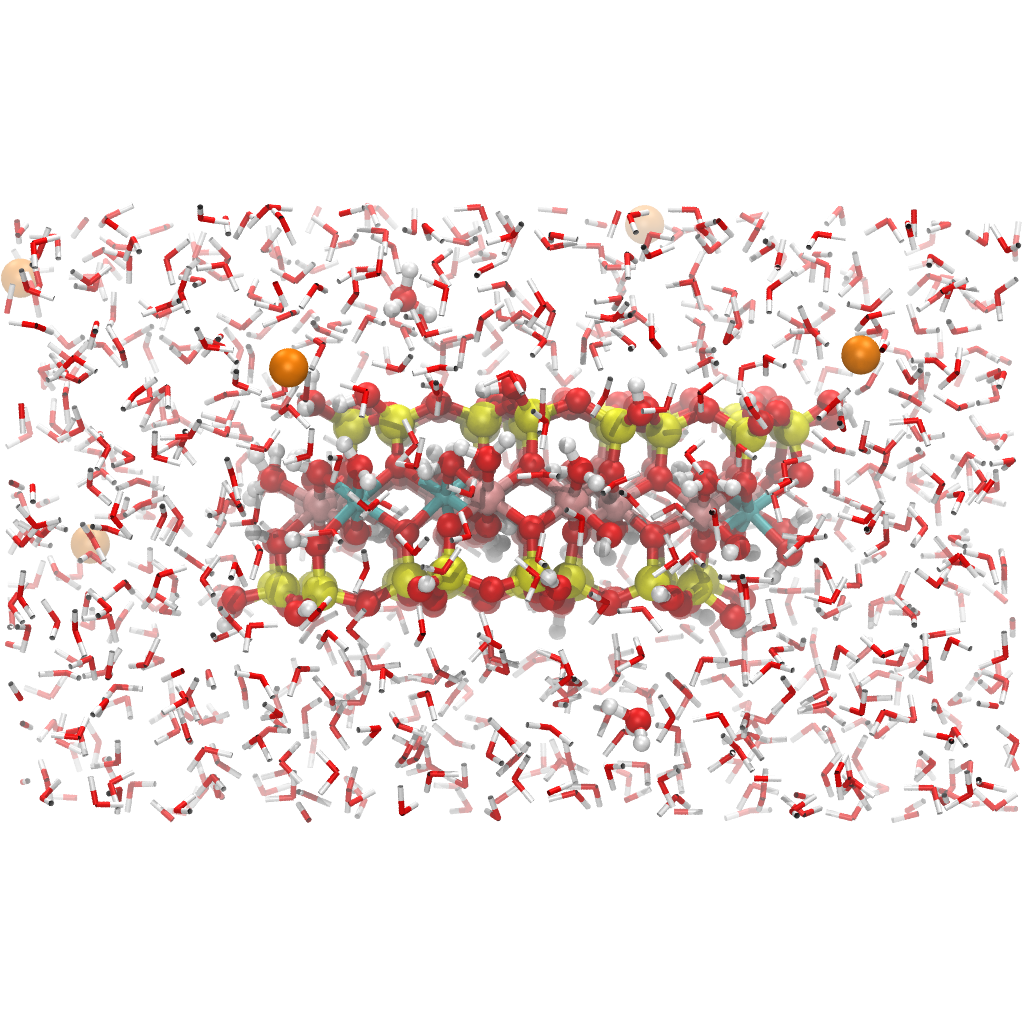} \\ \hline
    Mont.2–base solution interface & 40 & N$_{Mont.\ particle}$ = 1 \newline N$_{water}$ = 750 \newline N$_{Na^+}$ = 5 \newline N$_{OH^-}$ = 5 & \includegraphics[height=2.8cm]{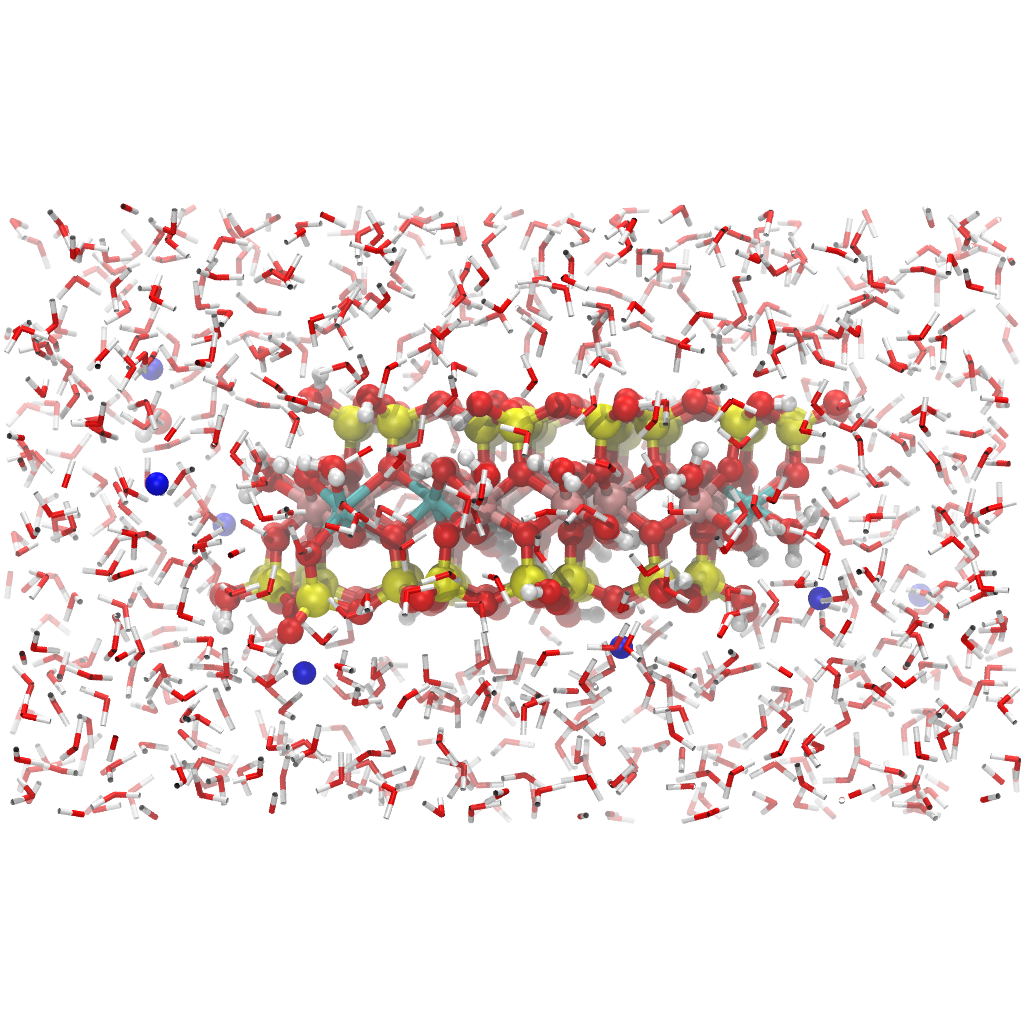} \\ \hline
\end{longtable}

\subsection{DFT Calculations for Reference Data}
To generate total energies and atomic forces for training the MLP, density functional theory (DFT) calculations were performed on all configurations in the dataset using the CP2K simulation package~\cite{doi:10.1063/5.0007045}. The Gaussian and plane wave (GPW) method was employed with a plane-wave cutoff of 1200~Ry to ensure high numerical accuracy~\cite{doi:10.1021/acs.jpclett.4c01030}. The exchange–correlation interaction was described by the revPBE functional~\cite{doi:10.1103/PhysRevLett.77.3865, doi:10.1103/PhysRevLett.80.890} combined with Grimme's D3 dispersion correction~\cite{doi:10.1063/1.3382344}. This functional was chosen due to its ability to provide a balanced and reliable description of both aqueous systems and clay minerals. In particular, it yields good accuracy in modeling the structure and dynamics of water~\cite{doi:10.1021/acs.jpclett.7b00391}, while also performing well in reproducing vibrational and structural properties of clay materials~\cite{doi:10.1063/5.0152361}. Electron–ion interactions were modeled using Goedecker–Teter–Hutter (GTH) pseudopotentials~\cite{doi:10.1103/PhysRevB.54.1703}, along with element-specific basis sets. A triple-$\zeta$ valence basis set with polarization (TZV2P-GTH) was used for H, O, Na, and Cl atoms, while a double-$\zeta$ valence basis set from the MOLOPT-SR family (DZVP-MOLOPT-SR-GTH) was used for Al, Mg, and Si. All calculations were conducted under periodic boundary conditions, with system dimensions defined according to the specific structure of each configuration. The energy and force outputs from these DFT calculations were used as the reference data for training of the MLP.

\subsection{Training of the Machine Learning Potential}
The MLP employed in this study was developed using the MACE framework~\cite{doi:10.48550/arXiv.2206.07697}, which combines message passing with high body-order equivariant features. This architecture has demonstrated convincing accuracy and transferability across a wide range of systems~\cite{doi:10.1063/5.0155322}. In this work, the MACE model was configured with two message-passing layers and four-body equivariant features, using a cutoff radius of 5~\AA.

To enable accurate modeling of the complex montmorillonite–aqueous interface, the potential was trained on a dataset composed of total energies and atomic forces calculated from DFT, as described in the previous section. The training set consisted of 500 configurations randomly sampled from the full DFT dataset, while an additional 80 configurations were reserved for testing. During training, 5\% of the data was used as validation set to monitor generalization performance.

\subsection{Modelcular Dynamics Simulations}
MLP-based MD simulations were carried out using the LAMMPS package under the NPT ensemble. A time step of 0.5~fs was employed. Temperature and pressure were maintained at 298~K and 1.01325~bar, respectively, using the Nosé–Hoover thermostat and barostat, with damping constants of 50~fs and 500~fs. To maintain structural stability, three central oxygen atoms in the clay particles were fixed throughout the simulations.

To systematically compare the behavior under different conditions, MD simulations were performed for all three montmorillonite nanoparticles (Mont.1, Mont.2, and Mont.3) in acidic, neutral, and basic aqueous environments, resulting in a total of nine systems. Each system was simulated for 1.2~ns, and configurations were sampled every 2~fs during the production runs. In the neutral systems, each nanoparticle was solvated with 750 water molecules. In acidic systems, five hydronium (H\textsubscript{3}O\textsuperscript{+}) ions and five chloride (Cl\textsuperscript{--}) ions were added, while in basic systems, five hydroxide (OH\textsuperscript{--}) ions and five sodium (Na\textsuperscript{+}) ions were included. Periodic boundary conditions were applied in all three dimensions for every system.
\clearpage

\section{Validation of Machine Learning Potentials}
\subsection{Validation on Interfacial Systems}
\begin{figure}[!htbp]
    \centering
    \includegraphics[width=\textwidth]{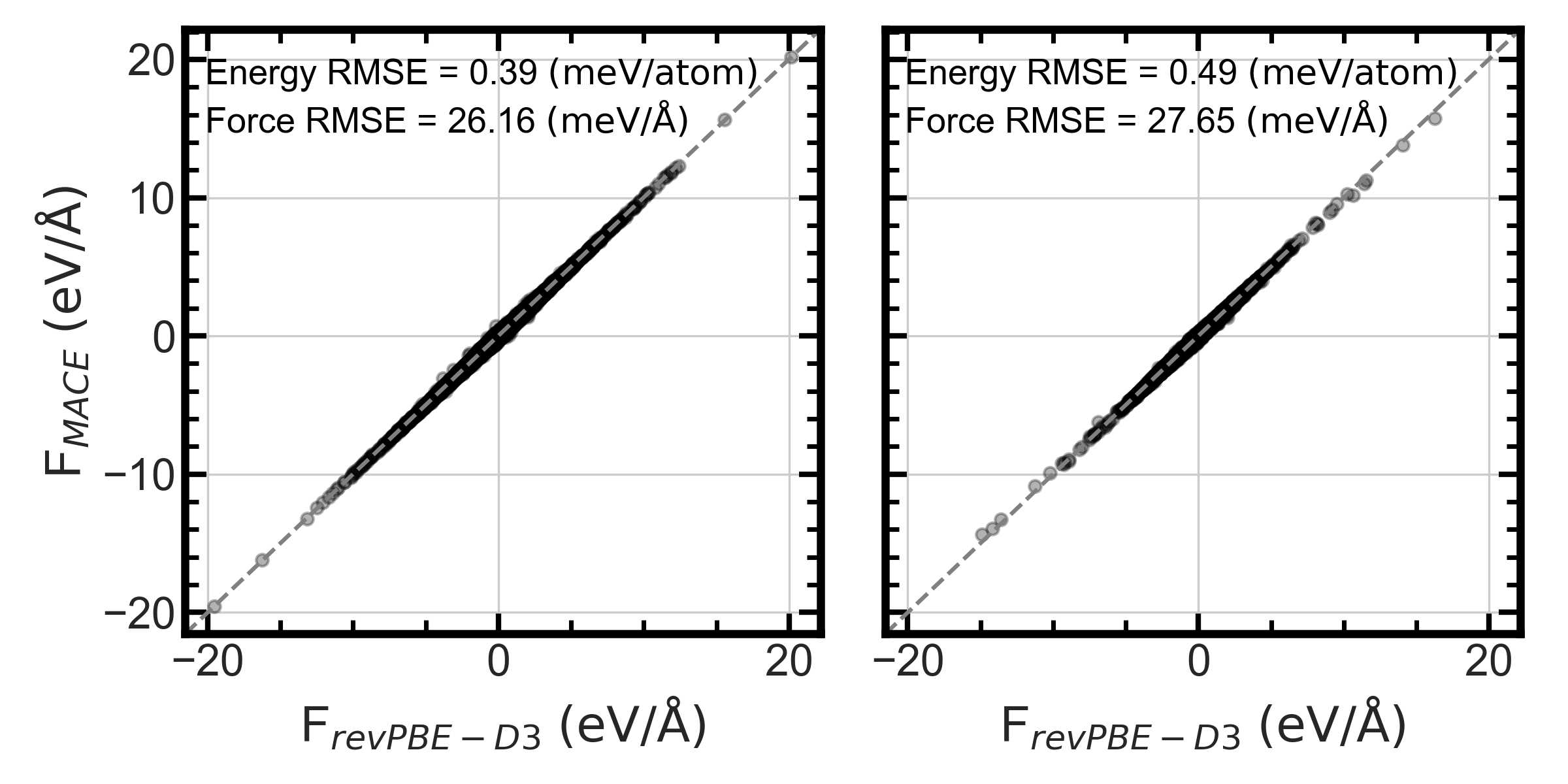}
    \caption{Correlation plots of atomic forces for the training (left) and test (right) datasets predicted by the MACE model compared to the DFT reference values. %
    The dashed grey line indicates a perfect correlation coefficient of 1.}
    \label{fig:FigureS2-1}
\end{figure}

Given the complexity of the modeled interface—comprising clay nanoparticles, aqueous solution, and the interaction region between them—we validate the MLP from three perspectives: its accuracy in predicting interfacial energies and forces, its structural reliability on clay systems, and its consistency with the structural features of bulk water. First, as shown in Figure~\ref{fig:FigureS2-1}, the root-mean-square errors (RMSEs) of energies and forces on the training and test datasets demonstrate that the MLP achieves high accuracy in reproducing the DFT reference data. 

\begin{figure}[!htbp]
    \centering
    \includegraphics[width=\textwidth]{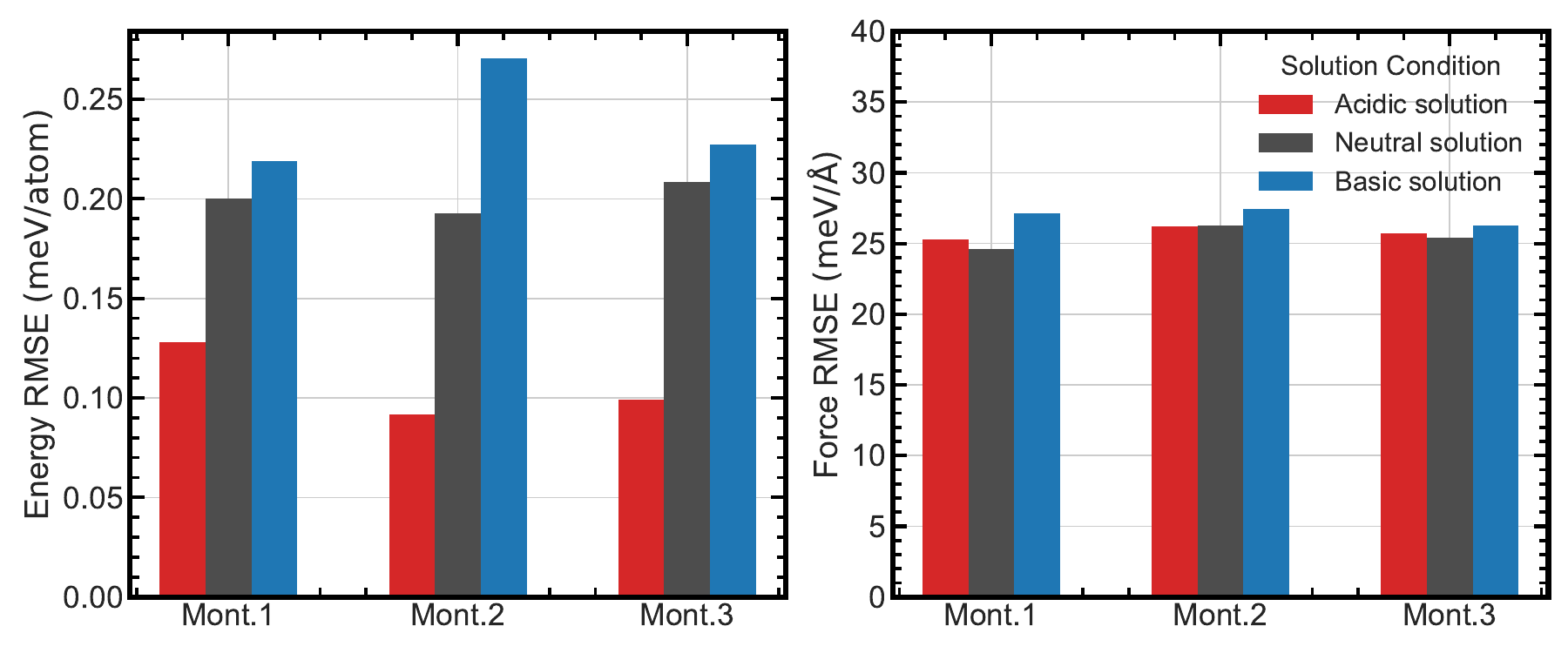}
    \caption{RMSE of energy and force predictions from the MLP compared to reference revPBE-D3 DFT results for nine interfacial systems.}
    \label{fig:FigureS2-2}
\end{figure}

To further validate the model in real simulation scenarios, we extracted 45 representative snapshots from the 1.2~ns production trajectory of each interfacial system (Mont.1, Mont.2, Mont.3, each combined with different aqueous solutions), resulting in a total of 405 interfacial configurations. DFT single-point calculations were performed on these snapshots to obtain reference energies and forces. As shown in Figure~\ref{fig:FigureS2-2}, the resulting RMSEs between the MLP predictions and the DFT references are less than 0.5~meV/atom for energies and 30~meV/\AA{} for forces. These results confirm that the MLP can reliably reproduce the energetics and interatomic interactions in complex montmorillonite–solution interfaces. 

It is worth noting that certain systems included in this evaluation were not part of the training set. For instance, the Mont.1 system surrounded by 750 water molecules (while only a 250-water-molecule configuration was included during training), as well as all configurations involving the Mont.3 clay structure, were excluded from the training dataset. Nonetheless, the MLP achieves excellent agreement with revPBE-D3 results across these systems, further demonstrating its transferability and robustness in capturing interfacial properties across diverse clay mineral structures.

\subsection{Validation on Pyrophyllite Supercell}
\begin{figure}[!htbp]
    \centering
    \includegraphics[width=\textwidth]{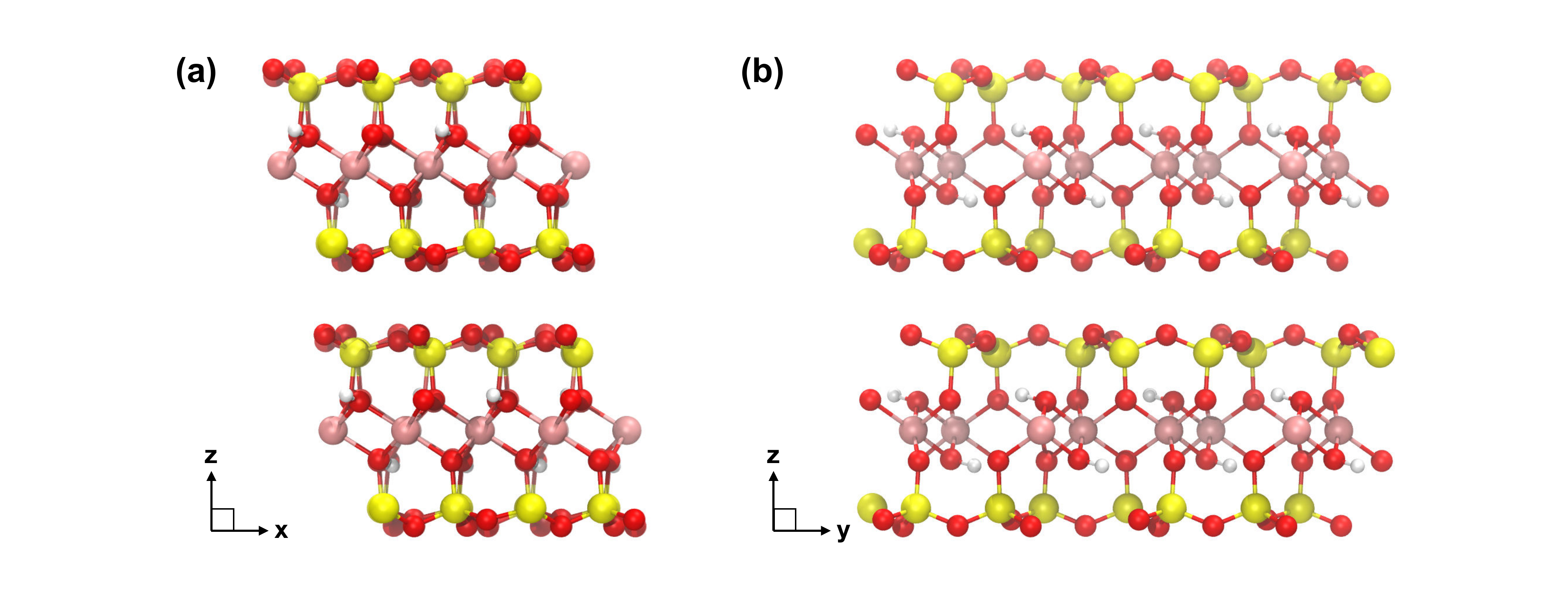}
    \caption{Snapshots of the pyrophyllite supercell (a) in the $xz$-plane and (b) in the $yz$-plane.}
    \label{fig:FigureS2-3}
\end{figure}

To assess the model's capability in reproducing clay mineral structures, we evaluated the relaxed cell parameters of a 2~$\times$~2~$\times$~2 pyrophyllite supercell (Figure~\ref{fig:FigureS2-3}). Pyrophyllite, like montmorillonite, is a dioctahedral 2:1 layered clay mineral with similar structural features, but it lacks isomorphic substitutions. As a result, its unit cell parameters are more uniform and well-characterized in experimental studies, whereas those of montmorillonite typically vary due to compositional heterogeneity and structural disorder.

In this study, both DFT and MLP optimizations were initiated from the same experimental pyrophyllite structure (see Ref.~\cite{doi:pyr_structure}). The DFT calculations employed the revPBE-D3 functional to directly perform full geometry and cell optimization. In contrast, the MLP-based workflow involved an initial 100~ps NVT molecular dynamics simulation at 300~K, followed by structural relaxation to determine the final equilibrium cell parameters. As summarized in Table~\ref{table:cell_para_bulk_pyr}, the MLP-predicted lattice parameters show less than 1\% deviation from the DFT-optimized values. Agreement with experimental data is also excellent. Minor discrepancies in certain bond lengths (greater than 1\%) are attributed to the limitations of the revPBE-D3 functional rather than the MLP. These results confirm that the trained MLP reliably reproduces the equilibrium geometry of layered clay minerals. Furthermore, the close agreement between MLP and DFT results for pyrophyllite---a structurally similar but chemically distinct system from montmorillonite---demonstrates the model's strong transferability across different types of dioctahedral 2:1 clay minerals. This highlights the potential of the MLP to generalize well to related clay systems beyond those explicitly included in the training set.

\begin{table}[!htbp]
    \centering
    \begin{threeparttable}
    \caption{Comparison of experimental, DFT, and MACE optimized structures, including cell parameters ($a$, $b$, $c$, $\alpha$, $\beta$, $\gamma$, volume) and representative bond lengths.}
    \label{table:cell_para_bulk_pyr}
    \renewcommand{\arraystretch}{1.2} 
    \begin{tabular}{c|c|c|ccc}
    \hline
    & EXP\textsuperscript{†} & revPBE - D3 & \multicolumn{3}{c}{MACE} \\ \cline{4-6}
    & Value & Value & Value & Error$_{DFT}$ ($\%$)  & Error$_{exp.}$ ($\%$) \\ \hline
    a (\AA) & 5.160 & 5.112 & 5.142 & 0.593 & 0.340 \\ 
    b (\AA) & 8.966 & 8.873 & 8.940 & 0.754 & 0.288 \\ 
    c (\AA) & 9.347 & 9.385 & 9.326 & 0.620 & 0.220 \\ 
    $\alpha$ (deg) & 91.180 & 90.831 & 91.068 & 0.260 & 0.123 \\ 
    $\beta$ (deg) & 100.460 & 100.931 & 100.550 & 0.377 & 0.090 \\ 
    $\gamma$ (deg) & 89.640 & 89.849 & 89.817 & 0.036 & 0.197 \\ 
    Volume (\AA$^3$) & 425.160 & 417.929 & 421.455 & 0.844 & 0.871 \\ 
    Si-O$_b$\textsuperscript{*} (\AA) & 1.612 & 1.625 & 1.626 & 0.016 & 0.850 \\ 
    Si–O$_a$\textsuperscript{*} (\AA) & 1.633 & 1.649 & 1.650 & 0.034 & 1.033 \\ 
    Al–O$_a$\textsuperscript{*} (\AA) & 1.915 & 1.930 & 1.933 & 0.150 & 0.923 \\ 
    Al–OH (\AA) & 1.889 & 1.887 & 1.890 & 0.125 & 0.039 \\ 
    O–H (\AA) & 0.935 & 0.966 & 0.965 & 0.061 & 3.261 \\ \hline
    \end{tabular}
    \begin{tablenotes}
    \footnotesize
    \item \textsuperscript{†} Experimental data are taken from Ref.~\cite{doi:pyr_structure}.
    \item \textsuperscript{*} O$_a$ and O$_b$ denote apical and basal oxygen atoms, respectively.
    \end{tablenotes}
    \end{threeparttable}
\end{table}

\subsection{Validation on Bulk Water}
\begin{figure}[!htbp]
    \centering
    \includegraphics[width=\textwidth]{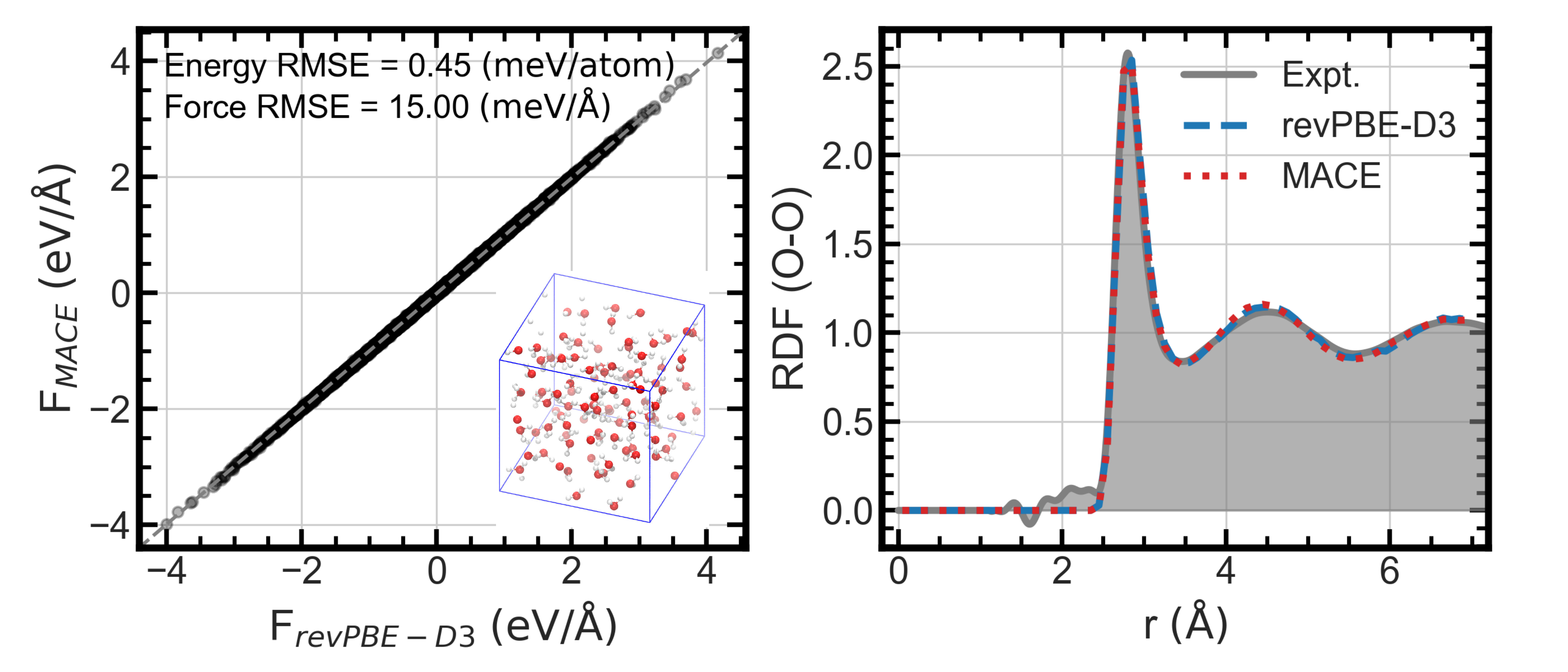}
    \caption{Correlation plot of atomic forces (left) and radial distribution function (RDF) comparison (right) for the bulk water system. %
    The left panel shows the correlation between forces predicted by the MACE model and those from DFT calculations. %
    The right panel compares the oxygen–oxygen RDFs computed from MACE (red dotted line), DFT (blue dashed line), and experimental measurements (gray solid line).
}
    \label{fig:FigureS2-4}
\end{figure}

To evaluate the MLP’s ability to capture the structural and energetic properties of liquid water, we performed additional tests on a bulk water system. A DFT-based \textit{ab initio} molecular dynamics (AIMD) simulation was carried out using CP2K on a system of 100 water molecules at 298~K in the NVT ensemble, producing a 10~ps trajectory. From this trajectory, 30 representative snapshots were selected, and both DFT and MLP energies and forces were evaluated. As shown in Figure~\ref{fig:FigureS2-4}, the RMSEs for energy and force are approximately 0.5~meV/atom and 15~meV/\AA{}, respectively, indicating a high level of agreement between the MLP and the DFT reference.

To further validate the structural accuracy of the MLP, we computed the oxygen–oxygen radial distribution function (RDF) from the MLP simulation and compared it with that from AIMD and experimental data~\cite{doi:10.1063/1.4790861}, as shown in Figure~\ref{fig:FigureS2-4}. The RDFs show excellent agreement across all three systems, indicating that the MLP can reliably reproduce the structural characteristics of liquid water.

\clearpage

\section{Acid--Base Reactivity of Different Nanoparticles}
\begin{figure}[!htbp]
    \centering
    \includegraphics[width=\textwidth]{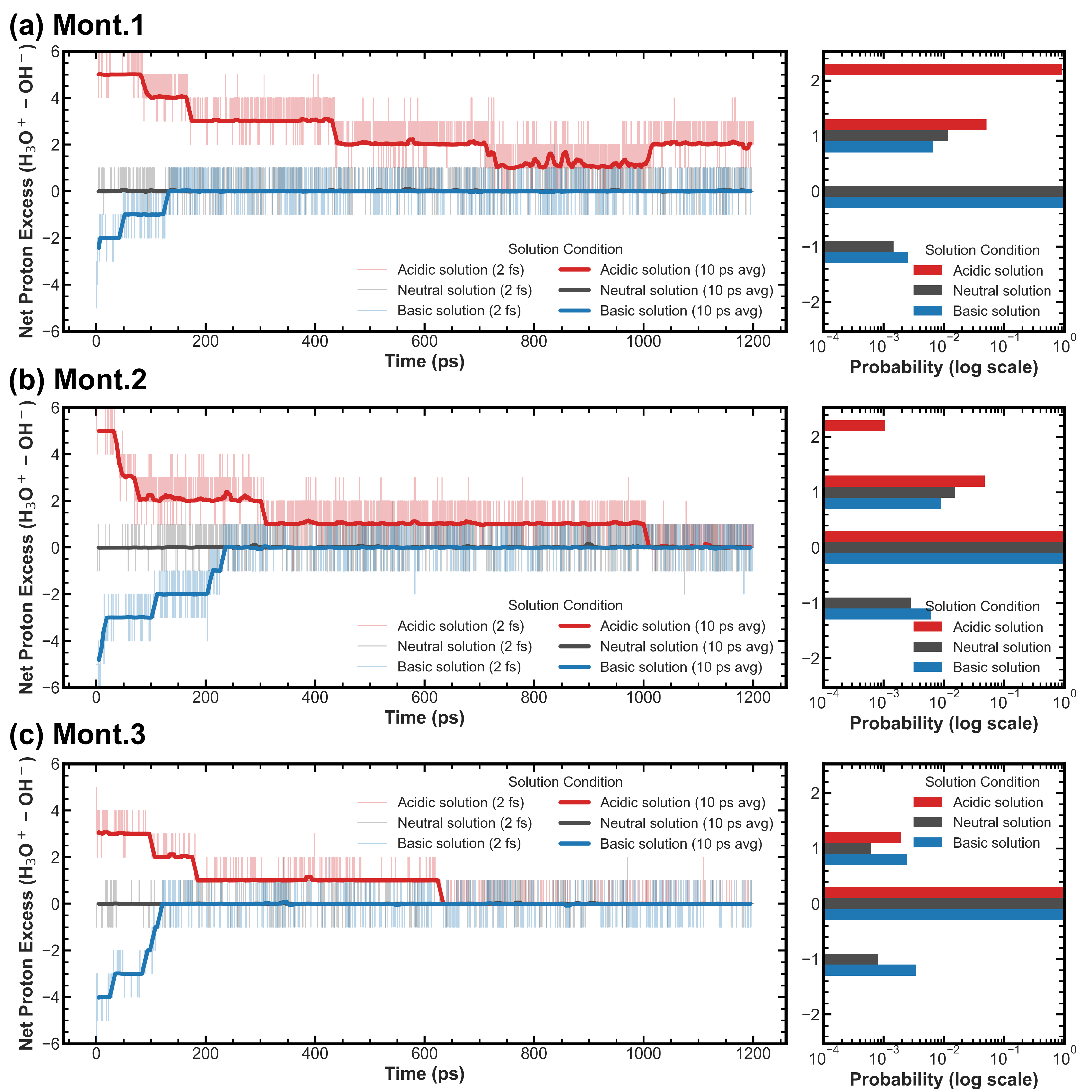}
    \caption{Time evolution and statistical distribution of the net proton excess in the aqueous phase for three montmorillonite systems. (a) Mont.1, (b) Mont.2, and (c) Mont.3. %
    In each panel, the left subfigure shows the time evolution of the net proton excess (defined as the number of hydronium ions minus hydroxide ions) over a 1200~ps molecular dynamics trajectory. Transparent curves represent raw data sampled every 2~fs, while solid lines correspond to time-averaged values over a 10~ps window. %
    The right subfigure presents the probability distribution of the net proton excess during the final 200~ps of the simulation. %
    Colors represent different solution pH conditions: red for acidic, black for neutral, and blue for basic environments.
    }
    \label{fig:FigureS3-1}
\end{figure}

In the main text, Mont.2 was selected as a representative structure to demonstrate the acid--base reactivity of montmorillonite nanoparticles in aqueous environments. Here, we extend the analysis to include Mont.1 and Mont.3, providing a comparative view of the interfacial protonation and deprotonation behaviors under acidic, neutral, and basic conditions.
Figure~\ref{fig:FigureS3-1} shows the time evolution of the net proton excess in the solution phase---defined as the difference between the number of hydronium ions and hydroxide ions (\ce{H_3O^+} $-$ \ce{OH^-})---over a 1.2~ns molecular dynamics trajectory, along with the corresponding distribution during the final 200~ps. All three montmorillonite systems exhibit amphoteric behavior, characterized by net proton uptake under acidic conditions and net proton release under basic conditions. Notably, deprotonation in basic solution proceeds much faster than protonation in acidic environments. This is reflected in the times required for five hydroxide ions to react in the basic solution: 182~ps for Mont.1, 235~ps for Mont.2, and 121~ps for Mont.3. In contrast, the times required for five hydronium ions to react in acidic solution are significantly longer: over 1.2~ns for Mont.1, 1019~ps for Mont.2, and 634~ps for Mont.3. Interestingly, in Mont.1 under acidic conditions, the net proton excess first decreases and then increases between 730~ps and 1000~ps, indicating a transient protonation and subsequent deprotonation of a \ce{-SiAlO^-} group.

\begin{figure}[!htbp]
    \centering
    \includegraphics[width=0.6\textwidth]{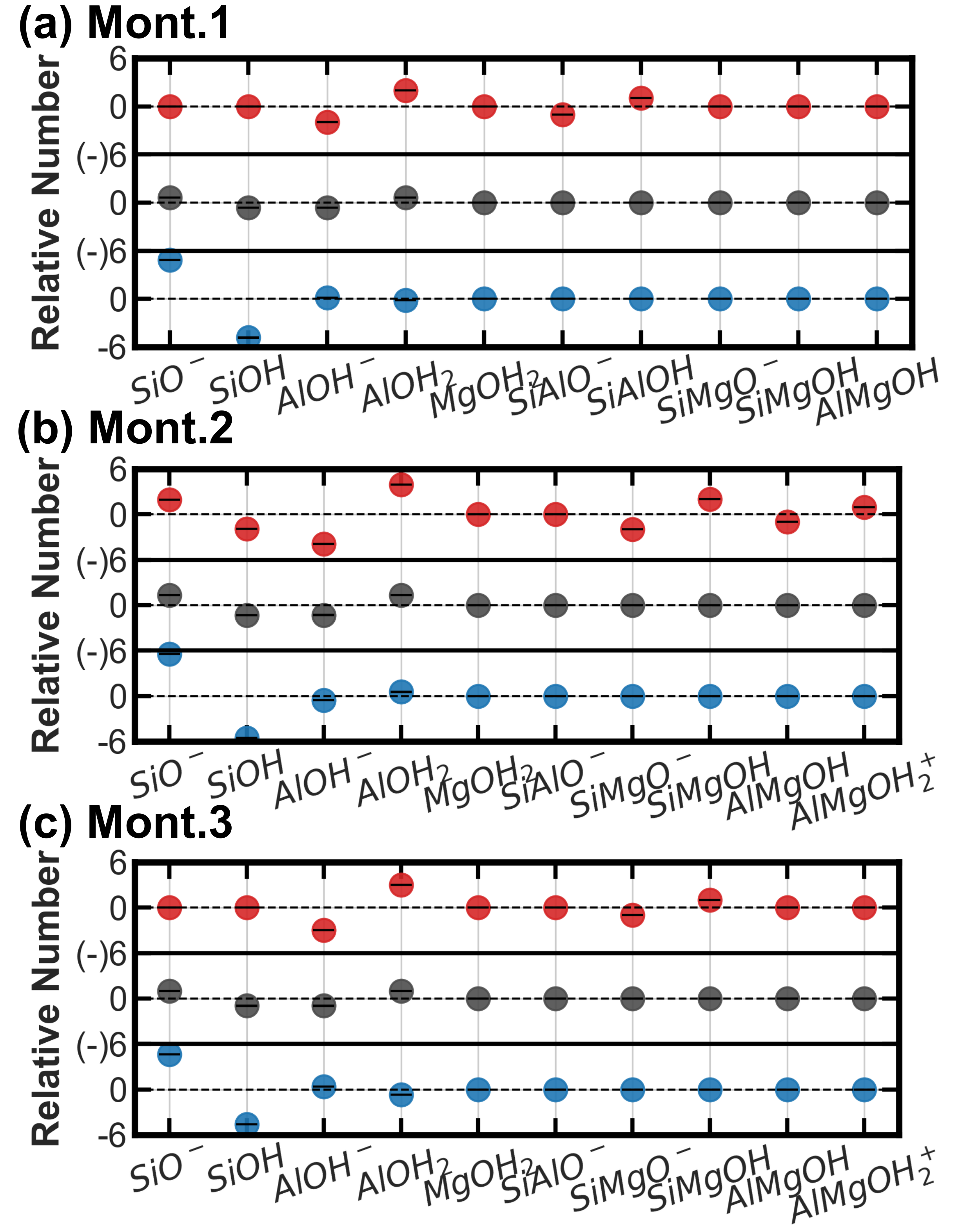}
    \caption{Relative abundance of surface functional groups on montmorillonite edges under different pH conditions. Panels (a), (b), and (c) correspond to Mont.1, Mont.2, and Mont.3, respectively. %
    Each point represents the relative population of specific surface functional groups during the final 200~ps of a 1.2~ns molecular dynamics trajectory, normalized by their initial populations. %
    Colors denote the solution pH: red for acidic, black for neutral, and blue for basic environments.}
    \label{fig:FigureS3-2}
\end{figure}

Figure~\ref{fig:FigureS3-2} presents the relative changes in surface functional group populations during the final 200~ps of the simulation, normalized to their initial values. In acidic solution, the patterns of protonation vary significantly among the three montmorillonite structures. For Mont.1, two \ce{-AlOH^-} groups are protonated to \ce{-AlOH_2}, and one \ce{-SiAlO^-} is protonated to \ce{-SiAlOH}, while two hydronium ions remain unreacted in solution. In Mont.2, two \ce{-SiMgO^-} groups are protonated to \ce{-SiMgOH}, one \ce{-AlMgOH} group becomes \ce{-AlMgOH_2^+}, and four \ce{-AlOH^-} groups are protonated to \ce{-AlOH_2}, including two that receive protons transferred from neighboring \ce{-SiOH} groups. Mont.3 shows protonation of four \ce{-AlOH^-} to \ce{-AlOH_2} and one \ce{-SiMgO^-} to \ce{-SiMgOH}.
In basic solution, the dominant deprotonation pathway involves \ce{-SiOH} groups. In Mont.2, protons are also transferred from \ce{-SiOH} to \ce{-AlOH^-}, resulting in the formation of \ce{-AlOH_2}; whereas in Mont.1 and Mont.3, \ce{-AlOH2} groups undergo partial deprotonation to \ce{-AlOH^-}. Under neutral conditions, proton transfer between \ce{-SiOH} and \ce{-AlOH^-} is observed across all three structures, highlighting the dynamic equilibrium of surface groups.

Overall, the spatial distribution of isomorphic substitutions in montmorillonite significantly influences both the number and type of reactive sites, particularly under acidic conditions. The observations from Mont.2 suggest that its specific isomorphic substitution pattern may promote either enhanced deprotonation of \ce{-SiOH} groups or increased proton affinity of neighboring \ce{-AlOH-} sites, thereby facilitating interfacial proton redistribution. These findings highlight both the common amphoteric nature of montmorillonite surfaces across different substitution patterns, as well as the subtle structural differences that modulate the extent and dynamics of acid--base reactivity at the interface.
\clearpage

\section{Proton Transfer Free Energy Landscape}
\subsection{Direct Proton Transfer}
To construct the proton transfer free energy landscape (PTFEL) for direct proton transfer (PT) between surface groups on the nanoparticle edge, each hydrogen atom was first assigned to its nearest oxygen atom based on interatomic distance. Hydrogen bonds were then identified using geometric criteria: the distance O$\mathrm{d}$--O$\mathrm{a}$ was required to be less than 3.5~\AA{}, and the angle O$\mathrm{a}$--O$\mathrm{d}$--H$_\mathrm{d}$ was constrained to be smaller than 30$^\circ$\cite{doi:10.1063/1.2431168}, where the subscripts d and a denote the hydrogen-bond donor and acceptor oxygen atoms, respectively.

\begin{figure}[!htbp]
    \centering
    \includegraphics[width=0.5\textwidth]{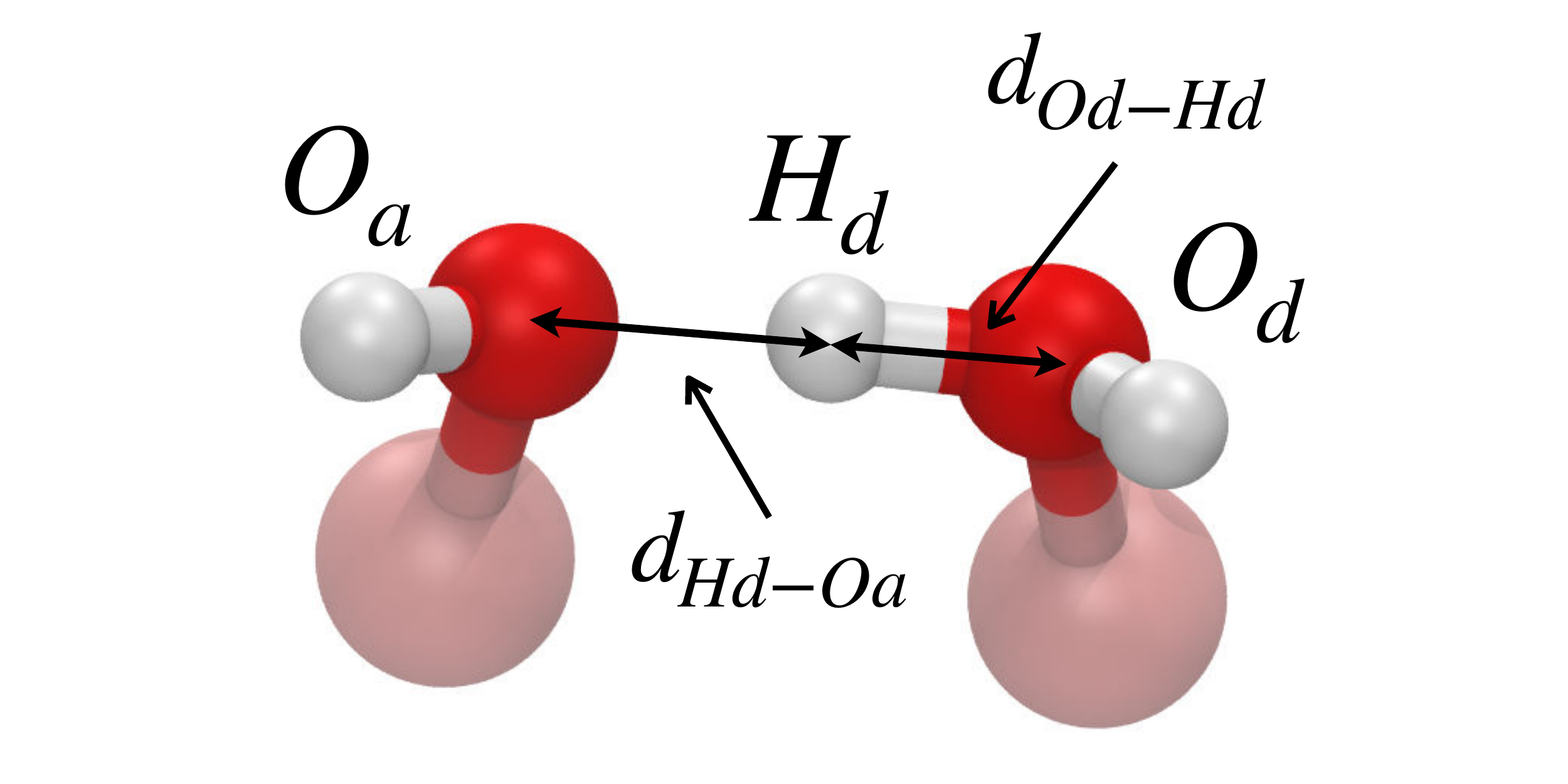}
    \caption{Schematic illustration of the variables used to describe a direct proton transfer reaction between an \(\mathrm{-AlOH^-}\) group and a neighboring \(\mathrm{-AlOH_2}\) group.}
    \label{fig:FigureS4-1}
\end{figure}

For each hydrogen-bond donor oxygen O$_\mathrm{d}$, all candidate acceptors O$_\mathrm{a}$ were evaluated, and a PT coordinate was defined as:
\[
\Delta = d_{\mathrm{Hd}-\mathrm{Oa}} - d_{\mathrm{Od}-\mathrm{Hd}}
\]
The pair with the minimum value of $\Delta$ for each donor oxygen was identified and denoted as $\delta = \Delta_\mathrm{min}$, representing the most probable PT event for that donor. These $\delta$ values were used as the reaction coordinates for constructing the free energy profiles.

Although $\delta$ is always computed as a positive quantity by definition, its sign was reassigned based on the direction of the underlying PT process. In the plotted PTFELs, negative values of $\delta$ correspond to the forward direction of the reaction, while positive values represent the reverse direction. The corresponding chemical equations are explicitly labeled in each PTFEL figure.

The values of $\delta$ were binned (with a bin width of 0.05 or 0.1~\AA), and the probability distribution $P(\delta)$ was estimated from histogram counts. The free energy profile was then computed as:
\[
\Delta F(\delta)/k_B T = -\ln P(\delta)
\]
where a small constant was added to avoid numerical divergence from $\ln(0)$. The resulting $\Delta F(\delta)/k_B T$ was shifted so that its minimum value is zero. 

\subsection{Solvent-assisted Proton Transfer}
To investigate the free energy landscape of solvent-assisted PT events, we identified representative configurations in which a water molecule bridges two reactive surface oxygen atoms (O$_1$ and O$_3$) on the nanoparticle edge. For each simulation frame, the nearest oxygen atom to both O$_1$ and O$_3$ was labeled O$_2$.

Following the same procedure as for direct PT, hydrogen atoms were assigned to their nearest oxygen atoms, and hydrogen bonds were identified using geometric criteria. Candidate hydrogens involved in the transfer were determined as follows: H$_1$ was defined as the hydrogen shared between O$_1$ and O$_2$, and H$_2$ as that between O$_2$ and O$_3$. Each hydrogen was chosen to minimize the distance mismatch between the two neighboring oxygens and further required to correspond to the shortest hydrogen bond for O$_1$ and O$_3$, respectively.

\begin{figure}[!htbp]
    \centering
    \includegraphics[width=0.5\textwidth]{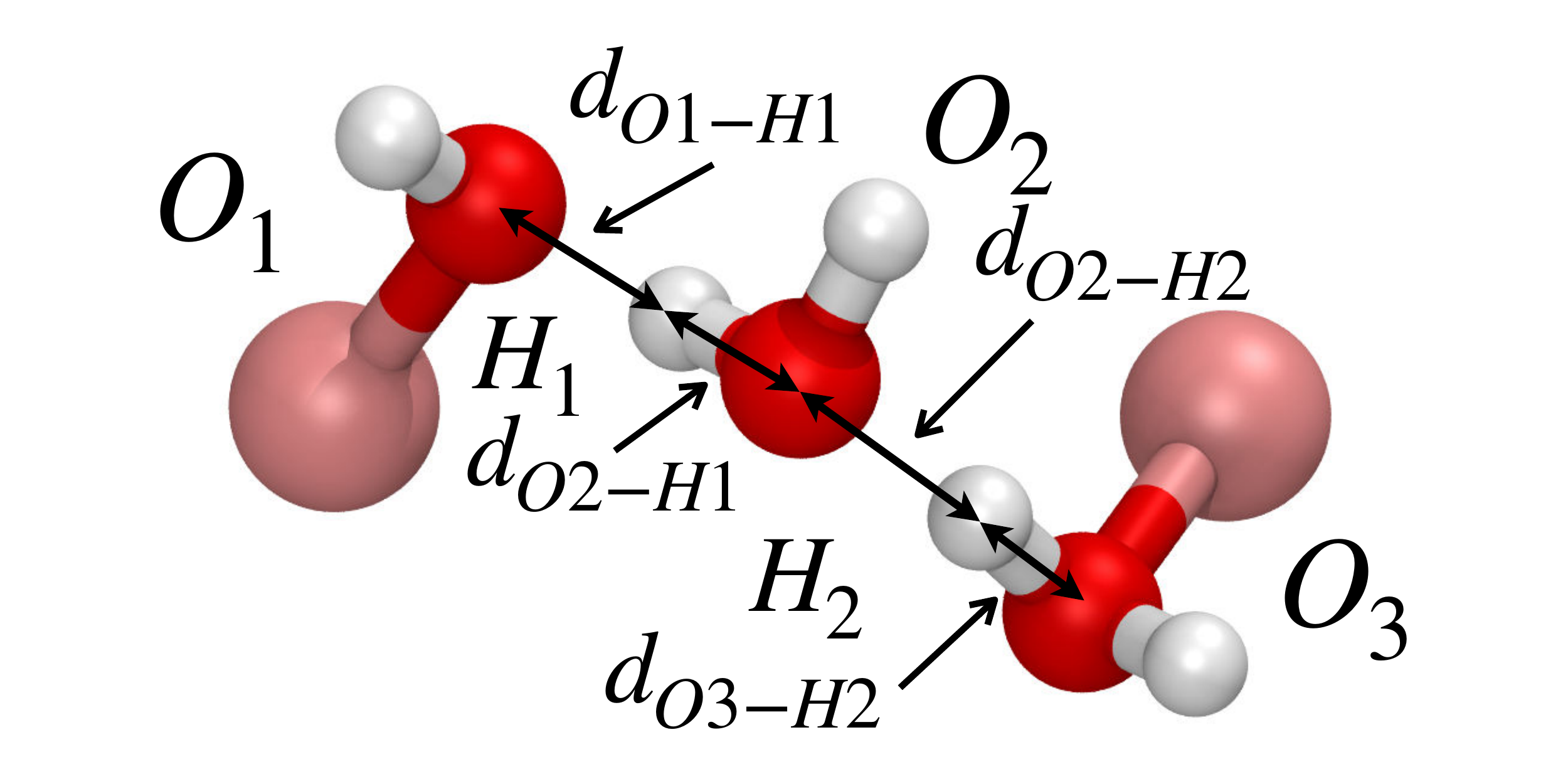}
    \caption{Schematic illustration of the collective variables used to describe a solvent-assisted proton transfer reaction between an \(\mathrm{-AlOH^-}\) group and a neighboring \(\mathrm{-AlOH_2}\) group. Two protons (H\textsubscript{1} and H\textsubscript{2}) bridge three oxygen atoms (O\textsubscript{1}, O\textsubscript{2}, and O\textsubscript{3}) via hydrogen bonding.}
    \label{fig:FigureS4-2}
\end{figure}

For accepted frames, donor–acceptor distances (O$_1$--H$_1$, O$_2$--H$_1$, O$_2$--H$_2$, O$_3$--H$_2$) were calculated (Figure~\ref{fig:FigureS4-2}). Two reaction coordinates were then defined to represent the proton positions relative to their donor and acceptor oxygen atoms:

\[
\xi_1 = d_{\mathrm{O2}-\mathrm{H1}} - d_{\mathrm{O1}-\mathrm{H1}}, \quad 
\xi_2 = d_{\mathrm{O2}-\mathrm{H2}} - d_{\mathrm{O3}-\mathrm{H2}}
\]

For one-dimensional analysis, the mean coordinate
\[
\xi = \frac{1}{2}(\xi_1 + \xi_2)
\]
was used, and its distribution was converted into a free energy profile by:
\[
\Delta F(\xi)/k_B T = - \ln P(\xi)
\]
The resulting profile was shifted so that its minimum energy was set to zero.

For two-dimensional free energy landscapes, joint histograms of $(\xi_1, \xi_2)$ were constructed, and the corresponding free energy surface was computed as:
\[
\Delta F(\xi_1, \xi_2)/k_B T = - \ln P(\xi_1, \xi_2)
\]
where $P(\xi_1, \xi_2)$ denotes the normalized two-dimensional probability density.
\clearpage

\subsection{Proton Transfer Events at the B Edge}

\begin{figure}[!htbp]
    \centering
    \includegraphics[width=\textwidth]{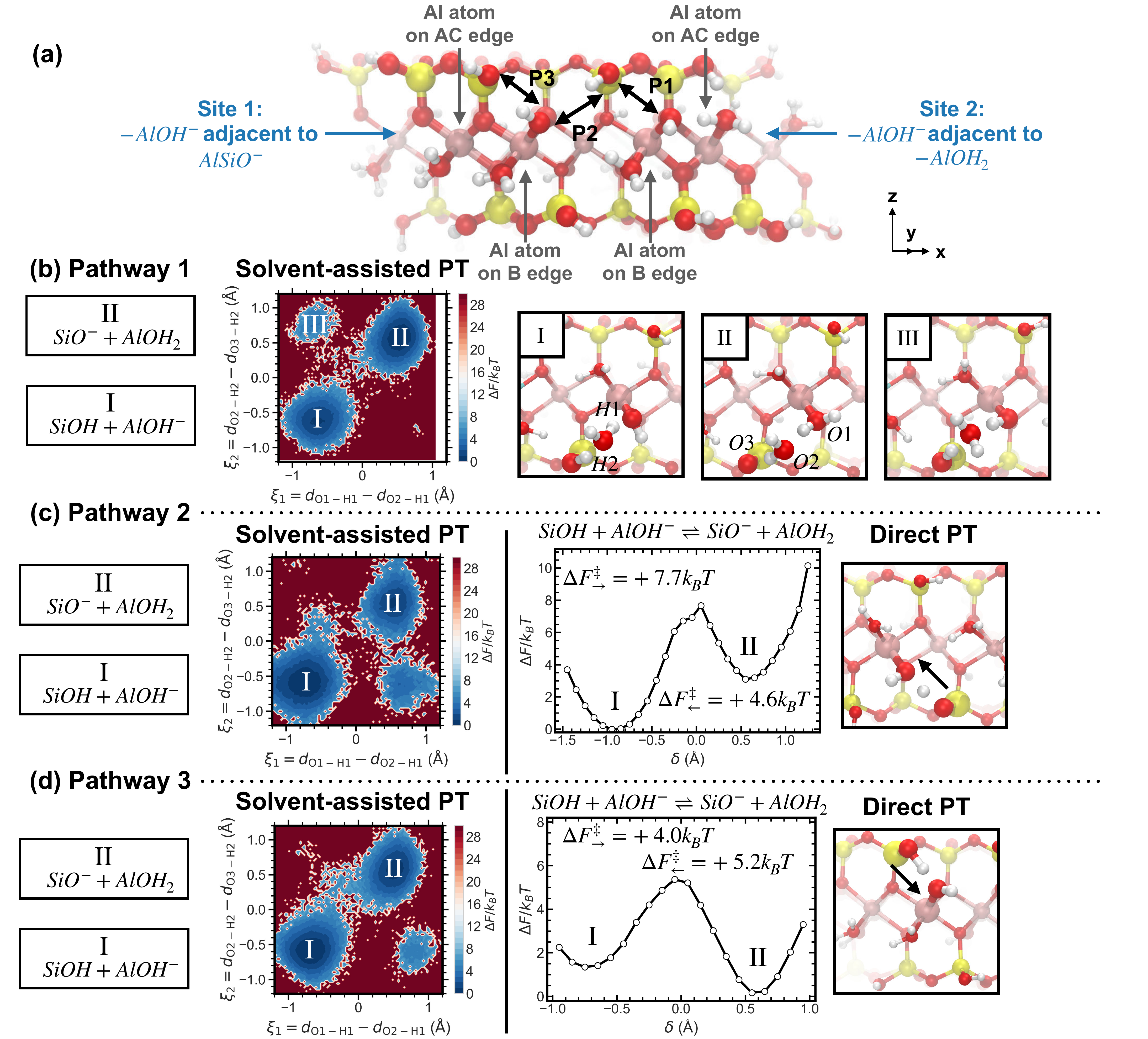}
    \caption{Proton transfer mechanisms at the B edge of montmorillonite under aqueous conditions. (a) Frontal view of the B edge. Black arrows indicate three distinct proton transfer pathways involving \(\mathrm{-SiOH}\) and \(\mathrm{-AlOH^-}\) groups. %
    (b) Pathway 1: Solvent-assisted proton transfer observed in neutral solution. Left: two-dimensional free energy surface projected onto collective variables \(\xi_1 = d_{\mathrm{O1}{-}\mathrm{H1}} - d_{\mathrm{O2}{-}\mathrm{H1}}\) and \(\xi_2 = d_{\mathrm{O2}{-}\mathrm{H2}} - d_{\mathrm{O3}{-}\mathrm{H2}}\), in units of \AA. Right: representative snapshots illustrating key intermediates along the reaction coordinate. %
    (c) Pathway 2: Proton transfer observed in basic solution. Left: free energy surface same as in (b), based on solvent-assisted dynamics. Right: free energy profile for the corresponding direct proton transfer pathway observed in neutral solution. %
    (d) Pathway 3: Proton transfer observed in basic solution. Left: solvent-assisted free energy surface, as defined above. Right: free energy profile for the direct proton transfer pathway under basic conditions.
}
    \label{fig:FigureS4-3}
\end{figure}

While the main text focused on a representative PT process at the B edge of montmorillonite in neutral solution, additional distinct pathways were identified across different interfacial environments. As shown in Figure.~\ref{fig:FigureS4-3}, we observed diverse PT events involving \ce{-SiOH} and \ce{-AlOH^-} groups under both neutral and basic conditions, further highlighting the dynamic reactivity of the B edge.

Pathway~1 involves a solvent-assisted proton transfer between a \ce{-SiOH} group and a neighboring \ce{-AlOH^-} site located at site~2 of the B edge, observed under neutral conditions. %
Although the relatively low number of transfer events in the trajectory limits the extraction of a well-defined one-dimensional free energy profile, the two-dimensional PTFEL in Figure~\ref{fig:FigureS4-3}a reveals that the free energy minimum corresponds to the protonated \ce{-AlOH_2} state, indicating a weak preference for protonation. %
For comparison, the direct proton transfer from an adjacent \ce{-AlOH_2} group at the AC edge (discussed in Section~B of the main text) has a lower barrier, making this solvent-assisted pathway energetically less favorable, but still mechanistically accessible within the simulated timescale.

Pathway~2 describes a direct PT between a \ce{-SiOH} group and an adjacent \ce{-AlOH^-} site of the B edge, observed in neutral solution. %
The forward barrier for \ce{-AlOH^-} protonation ($\Delta F^{\ddagger}_{\rightarrow} = 7.7\,k_BT$) is significantly larger than the reverse barrier ($\Delta F^{\ddagger}_{\leftarrow} = 4.7\,k_BT$), indicating that although the reaction can occur, it is less favorable and thus sampled less frequently (Figure~\ref{fig:FigureS4-3}b, right). %
In basic solution, a solvent-assisted pathway was identified at the same site. (Figure~\ref{fig:FigureS4-3}b, left) %
Due to the relatively low number of PT events and the influence of other stable, non-reactive configurations, a clear one-dimensional free energy profile could not be extracted from the two-dimensional surface, preventing direct comparison of activation barriers. %
Nevertheless, its comparable occurrence frequency indicates that this pathway can also proceed via solvent-assisted proton transfer.

Pathway~3 involves PT between a \ce{-SiOH} group located near the junction of the AC and B edges and a neighboring \ce{-AlOH^-} site at site~1. In basic solution, both direct and solvent-assisted pathways were observed to occur reversibly between the same pair of functional groups. (Figure~\ref{fig:FigureS4-3}c) Notably, the free energy profiles differ: the direct PT favors the formation of \ce{-AlOH_2} and \ce{-SiO^-}, while the solvent-assisted PT favors \ce{-AlOH^-} and \ce{-SiOH}. This observation highlights that not only the PT barriers but also the relative stability of the resulting protonation states can be influenced by the mechanism—direct versus solvent-assisted.

Together, these findings underscore the complexity of edge-site acid--base chemistry in montmorillonite and demonstrate that both local structure and interfacial environment play a critical role in determining reaction pathways and energetics.

\clearpage

\subsection{Proton Transfer Events at the AC Edge}
\begin{figure}[!htbp]
    \centering
    \includegraphics[width=0.7\textwidth]{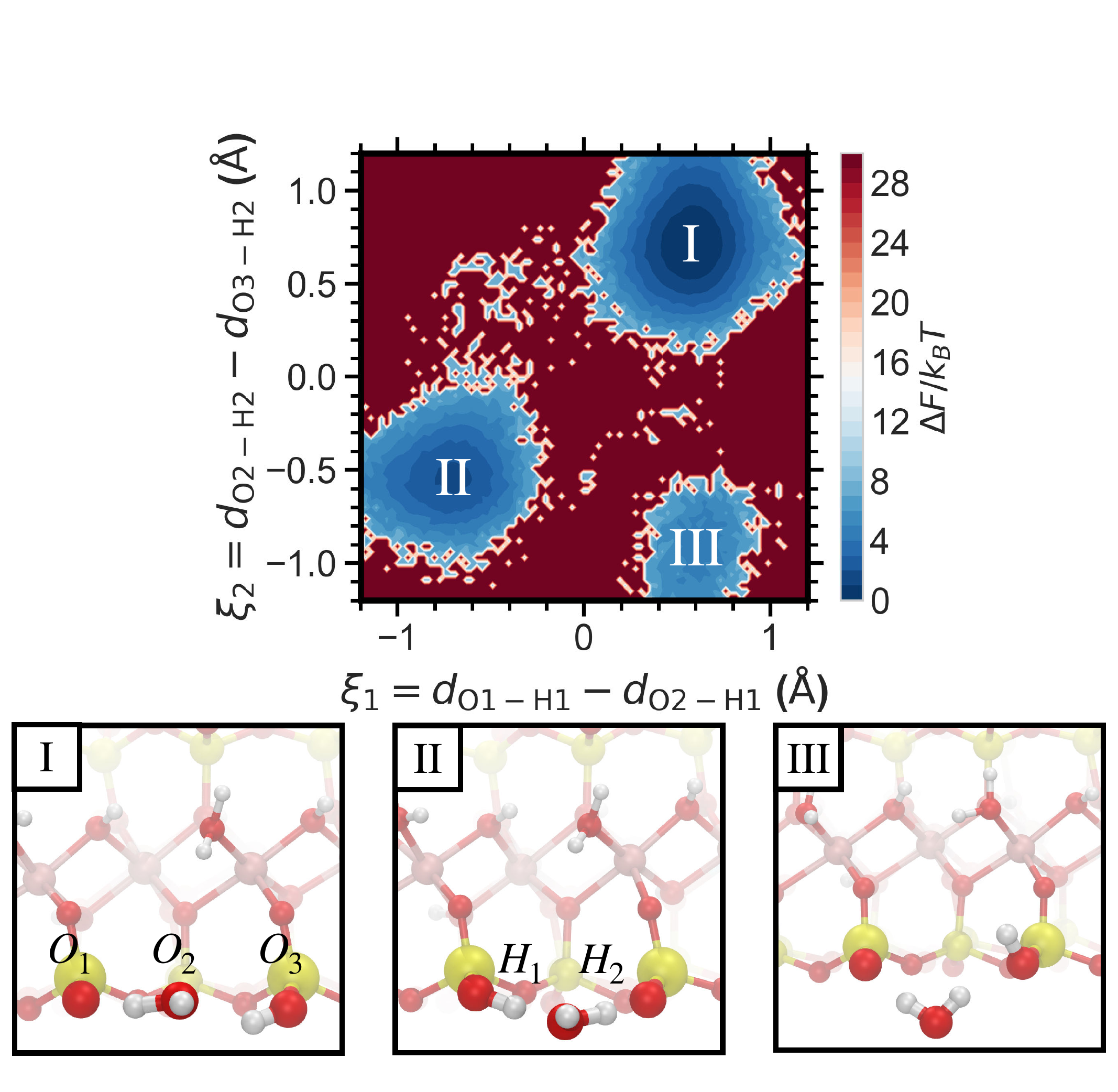}
    \caption{Proton transfer processes on the AC edge of the montmorillonite nanoparticle in basic solution. Two-dimensional free energy surface for a solvent-assisted proton transfer process between \(\mathrm{-SiO^-}\) and \(\mathrm{-SiOH}\) groups, plotted as a function of \(\xi_1\) and \(\xi_2\) (in \AA)
    }
    \label{fig:FigureS4-4}
\end{figure}

In this pathway, the initially deprotonated site is denoted as O$_1$. %
The 2D PTFEL in Figure~\ref{fig:FigureS4-4} shows no continuous low-energy route between stable states~I and~II, indicating an isolated, irreversible PT event rather than a reversible exchange. %
Trajectory inspection confirms that such transfers are rare over the 1.2~ns simulation, preventing reliable extraction of a one-dimensional free energy profile. %
A third minimum (state~III) corresponds to a stable, non-reactive solvation geometry where both hydrogen atoms of the bridging water molecule point toward surface groups, an orientation unfavorable for PT. %
These results demonstrate that solvent-bridged configurations do not universally promote PT; the hydrogen-bond network must be oriented appropriately to enable efficient transfer.
\clearpage

\subsection{Water-mediated Multi-step Proton Transfer Events}
In basic solution environments, we identified characteristic multi-step PT events at the AC edge, in which protons migrate between surface functional groups via bridging water molecules. As illustrated in Figure~\ref{fig:FigureS4-3}, two representative PT pathways were observed: (a) between adjacent \ce{-SiOH} and \ce{-SiO^-} groups, and (b) between a \ce{-SiOH} and a neighboring \ce{-AlOH^-} group. 
These PT events occur through water-mediated mechanisms involving transient configurations that resemble hydroxide-like species. The sequence of snapshots (Structures~I–IV) clearly demonstrates that the proton is not transferred as a single, localized particle in a continuous fashion. Instead, the process proceeds via a delocalized proton hole propagating through the hydrogen-bond network, characteristic of a Grotthuss-like transfer mechanism. This pathway underscores the dynamic role of water in facilitating long-range PT by stabilizing transient intermediate states and reducing the energetic barriers associated with proton relocation.

\begin{figure}[!htbp]
    \centering
    \includegraphics[width=\textwidth]{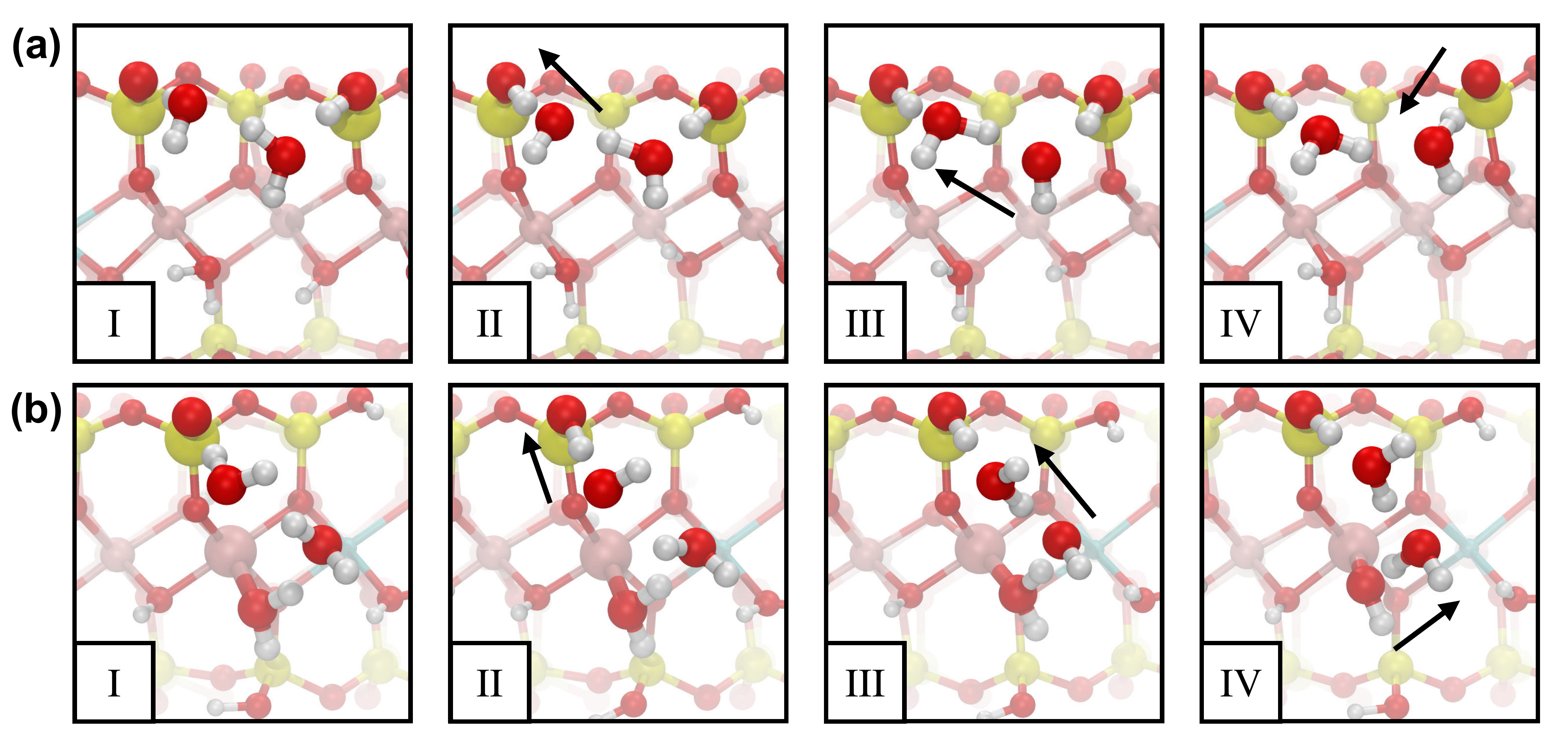}
    \caption{Representative snapshots illustrating water-mediated multi-step proton transfer events at the montmorillonite–water interface. %
    (a) Proton migration between \(\mathrm{-SiOH}\) and \(\mathrm{-SiO^-}\) groups. %
    (b) Proton migration between \(\mathrm{-SiOH}\) and \(\mathrm{-AlOH^-}\) groups. %
    Each panel consists of four snapshots (Structures I–IV), which depict the sequential transfer of a proton through bridging water molecules. Arrows indicate the direction of proton transfer.}
    \label{fig:FigureS4-5}
\end{figure}
\clearpage

\subsection{Isomorphic substitution influence in acidic and basic solution}
\begin{figure}[!htbp]
    \centering
    \includegraphics[width=\textwidth]{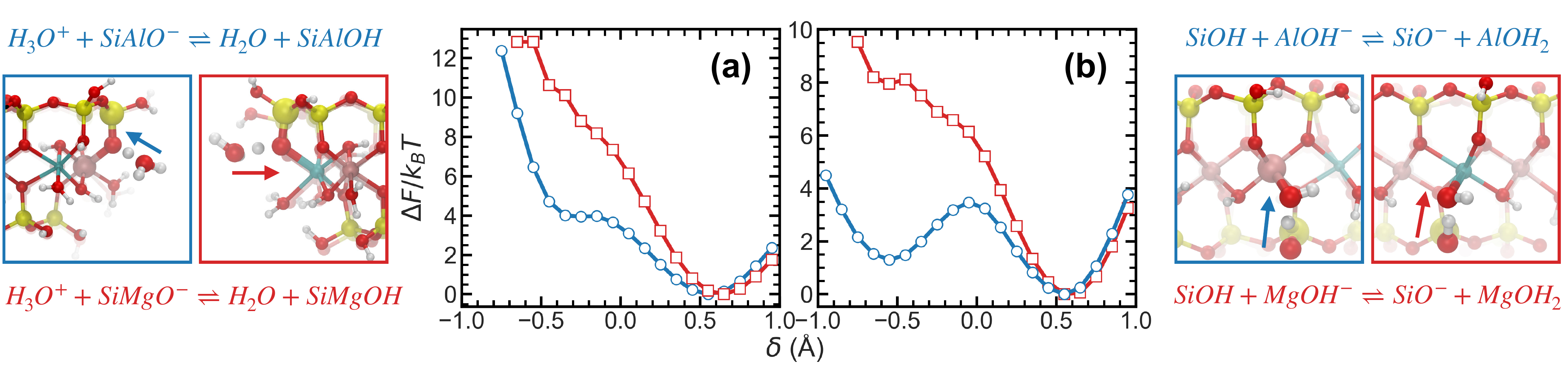}
    \caption{Effect of isomorphic substitution (Mg for Al) on the free energy profiles of direct proton transfer reactions in the montmorillonite nanoparticle system.
    Each panel compares two scenarios: red curves (highlighted by red frames) correspond to Mg-substituted sites, and blue curves (highlighted by blue frames) correspond to Al-only environments.
    For each case, the associated reaction equations and representative structures are shown alongside, with arrows indicating the direction of proton transfer.
     (a) Free energy profile for proton transfer between a hydronium ion (\ce{H3O+}) and either \ce{SiAlO^-} or \ce{SiMgO^-} in acidic solution. %
     (b) Free energy profile for a direct proton transfer on the AC edge, involving \ce{-SiOH} and either \ce{-AlOH^-} or \ce{-MgOH^-} groups that are not located near B-edge junctions, in basic solution.}
    \label{fig:FigureS4-6}
\end{figure}

In acidic solution, isomorphic substitution markedly enhances the proton affinity of edge sites that are otherwise less reactive. %
As shown in Fig.~\ref{fig:FigureS4-6}a, \ce{-SiAlO^-} and \ce{-SiMgO^-} groups both undergo protonation, with Mg substitution favoring stable retention of the acquired proton. %
Under basic conditions (Fig.~\ref{fig:FigureS4-6}b), \ce{-MgOH_2} groups display a thermodynamic preference for remaining protonated, even after transient deprotonation via reaction with hydroxide ions in solution. %
These observations are consistent with the trends discussed in the main text, highlighting the role of isomorphic substitution in modulating the acid--base reactivity of montmorillonite edge sites.

\clearpage

\subsection{Proton Transfer Events of \ce{-AlMgOH}}
In the main text, we demonstrated that under acidic conditions, \ce{-AlMgOH} groups located at the intersection of two AC edges can readily undergo protonation by hydronium ions, forming \ce{-AlMgOH_2^+}. In addition to this process, further analysis of trajectories under neutral conditions revealed transient PT events between neighboring \ce{-SiOH} and \ce{-AlMgOH} groups. These interactions involve a spontaneous attempt by \ce{-AlMgOH} to acquire a proton from the adjacent \ce{-SiOH}, thereby forming a transiently protonated \ce{-AlMgOH_2^+} species. However, this protonation is short-lived, with the proton quickly returning to the original donor site within approximately 100~fs.
To better understand the thermodynamic profile of this process, we computed the PTFEL for this direct PT. As shown in Figure~\ref{fig:FigureS4-4}, the protonation of \ce{-AlMgOH} by \ce{-SiOH} is associated with a significant free energy barrier ($\Delta F^{\ddagger}_{\rightarrow} = 7.5\,k_BT$), whereas the reverse reaction has a much lower barrier ($\Delta F^{\ddagger}_{\leftarrow} = 0.5\,k_BT$). These results suggest that while \ce{-AlMgOH} groups are capable of temporarily accepting a proton in neutral environments, the process is thermodynamically unfavorable and kinetically unstable, resulting in rapid reversion to the deprotonated state.

\begin{figure}[!htbp]
    \centering
    \includegraphics[width=\textwidth]{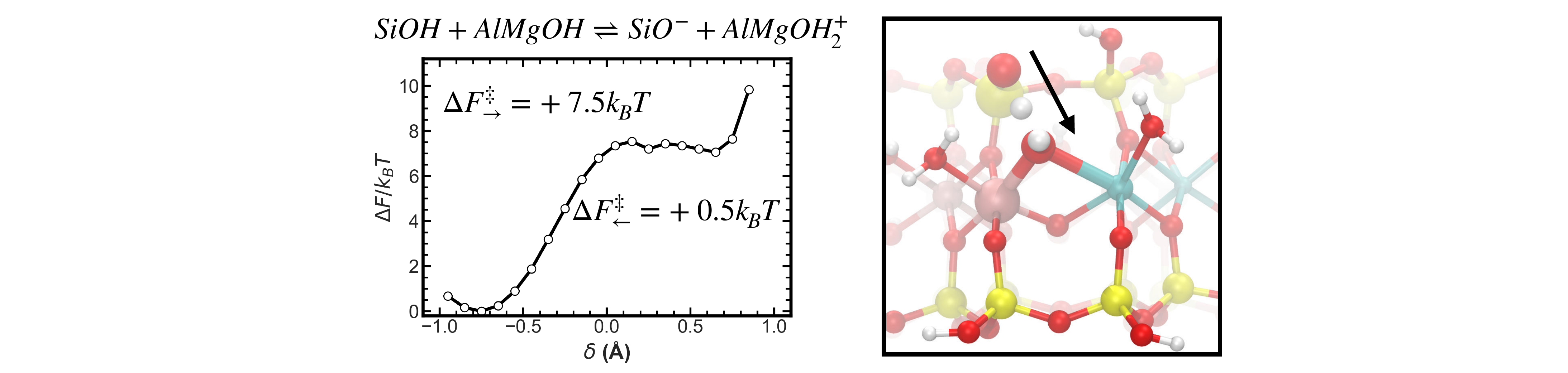}
    \caption{Free energy profile of a proton transfer reaction between a surface \(\mathrm{-SiOH}\) group and a neighboring \(\mathrm{-AlMgOH}\) site. The representative structure is shown adjacent to the energy curve, with an arrow indicating the direction of proton transfer.}
    \label{fig:FigureS4-7}
\end{figure}
\clearpage

\bibliographystyle{unsrt}
\bibliography{refs}